\documentclass[11pt]{article}

\usepackage[utf8]{inputenc}
\usepackage[T1]{fontenc}
\usepackage[english]{babel}
\usepackage{amsmath,amsthm,amssymb,mathtools}
\usepackage{aliascnt}
\usepackage[letterpaper,margin=1in]{geometry}
\usepackage{bbm}
\usepackage[hyphens]{url}
\usepackage{algorithm,algpseudocode}
\usepackage{booktabs,tabularx,array}
\usepackage{multicol} % Allows for multiple columns
\usepackage{multirow}
\usepackage[pagebackref,colorlinks]{hyperref}
\usepackage[capitalize,noabbrev]{cleveref}
\AtBeginDocument{%
  \DeclareFontShape{T1}{cmr}{m}{scit}{<->ssub*cmr/m/sc}{}%
}
\makeatletter
\providecommand{\theHALG@line}{}
\renewcommand{\theHALG@line}{\thealgorithm.\arabic{ALG@line}}
\makeatother

\newtheorem{theorem}{Theorem}[section]

\newaliascnt{lemma}{theorem}
\newtheorem{lemma}[lemma]{Lemma}
\aliascntresetthe{lemma}

\newaliascnt{proposition}{theorem}
\newtheorem{proposition}[proposition]{Proposition}
\aliascntresetthe{proposition}

\newaliascnt{corollary}{theorem}
\newtheorem{corollary}[corollary]{Corollary}
\aliascntresetthe{corollary}

\newaliascnt{remark}{theorem}
\newtheorem{remark}[remark]{Remark}
\aliascntresetthe{remark}

\crefname{lemma}{Lemma}{Lemmas}
\crefname{proposition}{Proposition}{Propositions}
\crefname{corollary}{Corollary}{Corollaries}
\crefname{remark}{Remark}{Remarks}

\DeclarePairedDelimiter\rbra{\lparen}{\rparen}
\DeclarePairedDelimiter\sbra{\lbrack}{\rbrack}
\DeclarePairedDelimiter\cbra{\{}{\}}
\DeclarePairedDelimiter\abs{\lvert}{\rvert}
\DeclarePairedDelimiter\norm{\lVert}{\rVert}
\DeclarePairedDelimiter\ceil{\lceil}{\rceil}
\DeclarePairedDelimiter\floor{\lfloor}{\rfloor}
\DeclarePairedDelimiter\ket{\lvert}{\rangle}
\DeclarePairedDelimiter\bra{\langle}{\rvert}
\newcommand{\ketbra}[2]{\ensuremath{\ket{#1}\bra{#2}}}
\DeclareMathOperator*{\E}{\mathbb{E}}
\DeclareMathOperator*{\Var}{Var}
\DeclareMathOperator{\tr}{tr}
\DeclareMathOperator{\rank}{rank}
\DeclareMathOperator{\supp}{supp}
\DeclareMathOperator{\sgn}{sgn}
\DeclareMathOperator{\diag}{diag}
\DeclareMathOperator{\Sym}{Sym}
\newcommand{\Id}{\mathbbm{I}}
\newcommand{\dtr}{\mathrm{d}_{\mathrm{tr}}}

\hypersetup{pdftitle={Optimal Trace Distance Estimation},
  pdfsubject={Sample complexity of trace distance estimation}}
\begin{document}
\title{Optimal Trace Distance Estimation}
\author{Ziyun Chen\footnote{Ziyun Chen is with the Paul G.\ Allen School of Computer Science \& Engineering, University of Washington 
(\href{mailto:ziyuncc@cs.washington.edu}{\nolinkurl{ziyuncc@cs.washington.edu}}).}
\and
Minbo Gao\footnote{Minbo Gao is with the Institute of Software, Chinese Academy of Sciences, and with University of Chinese Academy of Sciences (\href{mailto:gaomb@ios.ac.cn}{\nolinkurl{gaomb@ios.ac.cn}} or
\href{mailto:gmb17@tsinghua.org.cn}{\nolinkurl{gmb17@tsinghua.org.cn}}).}
\and
Qisheng Wang\footnote{Qisheng Wang is with the School of Computer Science, Shanghai Jiao Tong University (\href{mailto:QishengWang1994@gmail.com}{\nolinkurl{QishengWang1994@gmail.com}).}}}
\date{}
\maketitle

\begin{abstract}
We present a trace distance estimator with optimal sample complexity. 
Specifically, the trace
distance between two unknown $d$-dimensional quantum states can be estimated to within additive error
$\varepsilon$ using
\[
\Theta \! \left( \frac{d^2}{\varepsilon^2\log^2\rbra{d}\max\cbra{1,\varepsilon^2\log^2\rbra{d}}} \right)
\]
samples of each quantum state. This matches the lower bound of
\hyperlink{cite.Wan26}{Wang (2026)}. The estimator combines tomography and
polynomial approximation, with variance bounds from quantum Fourier analysis.
\end{abstract}

\newpage
\tableofcontents
\newpage

\section{Introduction}
\label{sec:introduction}

In quantum mechanics, trace distance is a metric on the space of density operators (cf.\ \cite{NC10,Wat18}).
The trace distance between two $d$-dimensional density operators $\rho$ and $\sigma$ is defined by
\[
\dtr\rbra{\rho,\sigma} = \frac{1}{2} \tr\rbra*{\abs{\rho-\sigma}}. 
\]
The Helstrom--Holevo bound \cite{Hel67,Hol73} reveals that the optimal probability of distinguishing two density operators $\rho$ and $\sigma$ is $\frac{1}{2}\rbra{1 + \dtr\rbra{\rho, \sigma}}$ (cf.\ \cite[Theorem 3.4]{Wat18}).
As a direct application, knowing the value of trace distance immediately gives this probability that is essential for discriminating density operators.
Practical \cite{ZRC19,ZR22} and variational \cite{LLSL21,CSZW22} approaches to trace distance estimation have been proposed for specific quantum systems. 

Trace distance estimation was first studied in \cite{Wat02} from the perspective of computational complexity, where they showed that trace distance estimation is $\mathsf{QSZK}$-hard. 
A simple trace distance estimator with sample complexity $O\rbra{\frac{d^2}{\varepsilon^2}}$, where $\varepsilon$ is the additive error, can be directly obtained by applying quantum state tomography to both input states \cite[Theorem 1.2 and Corollary 1.4]{OW16}. 
A central question is: 
\[
\textit{Can we estimate the trace distance with fewer samples?}
\]

\paragraph{Upper bounds.}
Positive results have been known for the low-rank case. 
A polynomial-time quantum algorithm for trace distance estimation between low-rank density operators was proposed in \cite{WGL+24}, confirming the Coles--Cerezo--Cincio conjecture \cite{CCC19} that low-rank trace distance estimation is in $\mathsf{BQP}$.
Later, a trace distance estimator was proposed in \cite{WZ24} when both density operators are of rank at most $r$, with sample complexity $O\rbra{\frac{r^2}{\varepsilon^5} \log^2\rbra{\frac{r}{\varepsilon}} \log^2\rbra{\frac{1}{\varepsilon}}}$, which was later improved to $O\rbra{\frac{r^2}{\varepsilon^4}\log^2\rbra{\frac{1}{\varepsilon}}}$ in \cite{CWZ25} and to $O\rbra{\frac{r^2}{\varepsilon^2}}$ in \cite{LT26}.
However, in the worst case where $r = d$, these bounds remain quadratic in $d$. 

Trace distance estimators have been known with significantly fewer samples in the very special case where at least one quantum state is pure (cf.\ \cite{Wan24,WZ26,FW25,LWY26,LW26}). 
In particular, in \cite{LWY26}, they showed that $\Theta\rbra{\frac{1}{\varepsilon^2}}$ samples are sufficient and necessary to estimate the trace distance when one quantum state is pure. 

\paragraph{Lower bounds.}
A sample complexity lower bound of $\Omega\rbra{\frac{d}{\varepsilon^2}}$ inherits from quantum mixedness testing \cite{OW21} and quantum state certification \cite{BOW19}.
Very recently, a near-quadratic lower bound of 
\[
\Omega\rbra*{\frac{d^2}{\varepsilon^2\log^2\rbra{d}\max\cbra{1,\varepsilon^2\log^2\rbra{d}}}}
\]
was established in \cite{Wan26}.

\subsection{Main result}

Given the aforementioned sample complexity upper and lower bounds for trace distance estimation, its tight bound still remains unclear. 
In this paper, we present an estimator for trace distance with sample complexity that is optimal only up to a constant factor.

\begin{theorem} \label{thm:main}
    Given two unknown $d$-dimensional quantum states $\rho$ and $\sigma$, there is a quantum algorithm that estimates the trace distance $\dtr\rbra{\rho, \sigma}$ to  within additive error $\varepsilon$ using 
    \[
    O\rbra*{\frac{d^2}{\varepsilon^2\log^2\rbra{d}\max\cbra{1,\varepsilon^2\log^2\rbra{d}}}}
    \]
    samples of $\rho$ and $\sigma$. 
\end{theorem}

The sample complexity upper bound given in \cref{thm:main} matches the lower bound established in \cite{Wan26}. 
In particular, for constant additive error $\varepsilon = \Theta\rbra{1}$, the sample complexity is $\Theta\rbra{\frac{d^2}{\log^4\rbra{d}}}$. 
Moreover, using the random dimension reduction technique in \cite{LT26} (for the upper bound) and combining the lower bound in \cite{Wan26}, we obtain a sample-optimal trace distance estimator in the low-rank case:
\begin{corollary} \label{coro:low-rank-intro}
    Given two unknown quantum states $\rho$ and $\sigma$ of rank at most $r$, it is sufficient and necessary to use
    \[
    \Theta\rbra*{\frac{r^2}{\varepsilon^2\log^2\rbra{r}\max\cbra{1,\varepsilon^2\log^2\rbra{r}}}}
    \]
    samples of $\rho$ and $\sigma$ to estimate the trace distance $\dtr\rbra{\rho, \sigma}$ to within additive error $\varepsilon$.
\end{corollary}

The sample complexity upper bound given in \cref{coro:low-rank-intro} improves the previous results in \cite{WZ24,CWZ25,LT26}, thereby settling the sample complexity of trace distance estimation. 

\begin{remark}[Comparison with optimal spectrum estimation \cite{BST26}]
    Recently, a quantum algorithm for spectrum estimation with optimal sample complexity was proposed in \cite{BST26}, which gives an estimator for the trace distance $\dtr\rbra{\rho, \Id/d}$ to the maximally mixed state with the same optimal sample complexity $\Theta\rbra{\frac{d^2}{\varepsilon^2\log^2\rbra{d}\max\cbra{1,\varepsilon^2\log^2\rbra{d}}}}$. 
    In comparison, our trace distance estimator applies to the case where both quantum states are unknown. 
    % Technically, our approach is based on polynomial approximation, which is quite different from \cite{BST26}. 
    Whereas \cite{BST26} recovers the spectrum by matching Chebyshev moments, our estimator directly estimates trace polynomials in the generally noncommuting matrices.
\end{remark}

We summarize the developments of trace distance estimation in \cref{tab:td}. 

\begin{table}[!htp]
    \centering
    \caption{Sample complexity of trace distance estimation.}
    \label{tab:td}
    \begin{tabular}{ccc}
    \toprule
    Complexity Type & Sample Complexity & References \\
    \midrule
    \multirow{13}{*}{Upper Bounds} & $O\rbra*{\dfrac{d^2}{\varepsilon^2}}$ & Implied by \cite{OW16} \\
    \addlinespace
    & $O\rbra*{\dfrac{r^2}{\varepsilon^5} \log^2\rbra*{\dfrac{r}{\varepsilon}} \log^2\rbra*{\dfrac{1}{\varepsilon}}}$ & \cite{WZ24} \\
    \addlinespace
    & $O\rbra*{\dfrac{r^2}{\varepsilon^4}\log^2\rbra*{\dfrac{1}{\varepsilon}}}$ & \cite{CWZ25} \\
    \addlinespace
    & $O\rbra*{\dfrac{r^2}{\varepsilon^2}}$ & \cite{LT26} \\
    \addlinespace
    & $O\rbra*{\dfrac{d^2}{\varepsilon^2\log^2\rbra{d}\max\cbra{1,\varepsilon^2\log^2\rbra{d}}}}$ when $\sigma = \mathbb{I}/d$ & Implied by \cite{BST26} \\
    \addlinespace 
    & $O\rbra*{\dfrac{r^2}{\varepsilon^2\log^2\rbra{r}\max\cbra{1,\varepsilon^2\log^2\rbra{r}}}}$ & This Work \\
    \midrule
    \multirow{3}{*}{Lower Bounds} & $\Omega\rbra*{\dfrac{d}{\varepsilon^2}}$ & \cite{OW21} \\
    \addlinespace
    & $\Omega\rbra*{ \dfrac{d^2}{\varepsilon^2\log^2\rbra{d}\max\cbra{1,\varepsilon^2\log^2\rbra{d}}} }$ & \cite{Wan26} \\
    \bottomrule
    \end{tabular}
\end{table}

\subsection{Technique overview}
\label{sec:overview}

\paragraph{Polynomial estimation.}
Set
\begin{equation}
S_+=\frac{\rho+\sigma}{2},\qquad S_-=\frac{\rho-\sigma}{2}.
\label{eq:average-difference}
\end{equation}
Then $\dtr\rbra{\rho,\sigma}=\tr\rbra{\abs{S_-}}$ and
$-S_+\preceq S_-\preceq S_+$. We approximate $\tr\rbra{\abs{S_-}}$ by a trace
polynomial in $S_+,S_-$ and estimate it without bias. Choose each sample
uniformly from $\rho,\sigma$ and label it $+1,-1$, respectively. A cyclic
permutation on $k$ samples, multiplied by their labels, has expectation
$\tr\rbra{S_-^k}$. Omitting a label replaces the corresponding factor by
$S_+$. Thus we can obtain unbiased estimates of traces of products of $S_+$ and $S_-$.
Averaging over subsets gives a quantum $U$-statistic~\cite{EAO02,GB10}.

For a trace polynomial $f$, let $G_{f,n}$ denote the sum of these averaged
kernels on $n$ labelled samples. Its measurement outcome $\widehat f_n$ satisfies
$\E\sbra{\widehat f_n}=f\rbra{S_+,S_-}$.
We measure $G_{f,n}$ and bound its variance by analyzing all polynomial
terms together, without assuming that $S_+$ and $S_-$ commute.

\paragraph{Tomography and a correction term.}
The variance bound depends on the operator norm of $S_+$. We first use
$O\rbra{d\log^2\rbra{d}}$ samples to estimate $S_+$ by density operator tomography.
From this estimate, we obtain a projector $Q$ such that, with constant
success probability, $\norm{QS_+Q}_\infty\leq1/\log\rbra{d}$ and
$\rank\rbra{\Id-Q}\leq2\log\rbra{d}$.
Using fresh samples, we estimate $\tr\rbra{\abs{QS_-Q}}$ and the
correction $\tr\rbra{\abs{S_-}}-\tr\rbra{\abs{QS_-Q}}$.

For the latter term, take an even approximation polynomial $q_K$ for absolute
value whose derivative satisfies
\[
\int_{-1}^1\abs{\sgn\rbra{x}-q_K'\rbra{x}}\,\mathrm d x=O\rbra{1/K}.
\]
\Cref{lem:trace-errors} shows that the error in the
correction term is at most $O\rbra{\rank\rbra{S_--QS_-Q}/K}$. Since
$S_--QS_-Q=\rbra{\Id-Q}S_-+QS_-\rbra{\Id-Q}$, the rank is at most $2\rank\rbra{\Id-Q}$. Thus degree
$K=O\rbra{\log\rbra{d}/\varepsilon}$ suffices, including the contribution
of the off-diagonal blocks of $S_-$.

\paragraph{Choosing the approximation polynomial.}
When $\varepsilon\log\rbra{d}\leq1$, a degree-$O\rbra{d/\varepsilon}$ polynomial
approximating $\abs{x}$ gives bias $O\rbra{\varepsilon}$ after summing over the
compressed eigenvalues. When $\varepsilon\log\rbra{d}>1$, we use polynomial
filters $R_b$ at dyadic scales $b$, with $0\preceq R_b\rbra{S_+}\preceq\Id$
and $\norm{R_b\rbra{S_+}S_-R_b\rbra{S_+}}_\infty=O\rbra{b}$.
The trace-norm differences between adjacent scales telescope.
Approximating them with scale-dependent degrees gives a trace polynomial
$F_{d,\varepsilon}$ of degree $O\rbra{\sqrt d/\varepsilon}$ and error at most $\varepsilon$,
which we estimate as above.

\paragraph{Bounding the variance.}
We bound the Fourier coefficients of $G_{f,n}$ with respect to the
labelled product density operator $\tau^{\otimes n}$. For full-rank $\tau$,
Parseval's identity and \cref{lem:fourier} give
\[
\Var\sbra*{\widehat f_n}
=\sum_{j=1}^{\deg\rbra{f}}W_j\rbra{G_{f,n}},
\qquad
W_j\rbra{G_{f,n}}\leq\binom{n}{j}^{-1}\norm{B_j}_{\tau,j}^2,
\]
where $W_j$ is the sum of squared Fourier coefficients of degree $j$,
and $B_j$ is an effective kernel on $j$ pairs. The resulting variance
bounds extend to singular density operators by continuity. For a univariate polynomial, the
kernel formula of \cite[Lemma~4.2]{CGW26} reduces the problem to scalar
derivative bounds. We extend that formula to signed labels and compression.

For the multiscale polynomial, we instead expand around $\rbra{S_+,S_-}$ and
represent each Taylor coefficient by ordered matrix products. Bounds on
separated divided differences give a factor $1/\rbra{2j}!$ in the order-$j$
coefficient. \Cref{lem:cycles} gives a further factor $1/\rbra{j-1}!$ in
the Fourier weight bound. Combining these bounds yields exponential factors
with leading exponents of order $K/\sqrt n$ in the first regime and
$\sqrt d/\rbra{\varepsilon n^{1/4}}$ in the second. The sample sizes in
\eqref{eq:sample-sizes} give standard deviation $O\rbra{\varepsilon}$.
Combining this with the bias bounds proves \cref{thm:main}.

\paragraph{Organization.}
\Cref{sec:preliminaries} introduces the notation, tomography,
approximation polynomials, and polynomial observables.
\Cref{sec:estimator} gives the estimator and its bias and variance bounds.
\Cref{sec:variance} develops the quantum Fourier analysis and proves the
variance bound, and \cref{sec:main-proof} completes the proof of \cref{thm:main}.
The approximation and Taylor-coefficient estimates are proved in the appendices.

\section{Preliminaries}
\label{sec:preliminaries}

\subsection{Notation and tomography}

For a positive integer $n$, let $[n]=\cbra{1,\ldots,n}$, and let $\log$ denote
the natural logarithm. A density operator on $\mathbb C^d$ is a positive semidefinite
matrix of trace one. We write $X\preceq Y$ when $Y-X$ is positive semidefinite
and $\Id$ for the identity on the space in question.
For a matrix $X$ with singular values $s_i$, its Schatten $p$-norm is
\[
\norm{X}_p=\rbra*{\sum_i s_i^p}^{1/p}\quad\rbra{1\leq p<\infty},
\qquad \norm{X}_\infty=\max_i s_i.
\]
The cases $p=1,2,\infty$ give the trace, Hilbert--Schmidt, and operator
norms, respectively.

For Hermitian $X=\sum_i\lambda_i\ketbra{v_i}{v_i}$, functional calculus means
$f\rbra{X}=\sum_i f\rbra{\lambda_i}\ketbra{v_i}{v_i}$. In particular, this defines
$\abs{X}$ and $\sgn\rbra{X}$, with $\sgn\rbra{0}=0$.
For an interval $J$, $\mathbf1_J\rbra{X}$ denotes the spectral projector.
All implicit constants in asymptotic notation are universal. Constants
denoted by $C$ may differ between bounds.

\begin{proposition}[Order and norm bounds]
\label{prop:source-order}
Let $\rho,\sigma$ be density operators on $\mathbb C^d$, and set
$S_+=(\rho+\sigma)/2$ and $S_-=(\rho-\sigma)/2$. Then
\begin{equation}
S_+\succeq0,\qquad\tr\rbra*{S_+}=1,\qquad -S_+\preceq S_-\preceq S_+,
\qquad\norm{S_-}_\infty\leq\frac12.
\label{eq:source-order}
\end{equation}
For every orthogonal projector $Q$, we also have $QS_+Q\succeq0$,
$-QS_+Q\preceq QS_-Q\preceq QS_+Q$, and $\tr\rbra{QS_+Q}\leq1$.

More generally, for any Hermitian pair $S_+,S_-$ with $S_+\succeq0$ and
$-S_+\preceq S_-\preceq S_+$, there is a Hermitian matrix $J$ such that
\begin{equation}
S_-=S_+^{1/2}JS_+^{1/2},\qquad \norm{J}_\infty\leq1.
\label{eq:factorization}
\end{equation}
Consequently,
$\norm{S_-}_p\leq\norm{S_+}_p$ for $1\leq p\leq\infty$.
\end{proposition}

\begin{proof}
The identities $S_++S_-=\rho$ and $S_+-S_-=\sigma$ give the order bounds.
Since $0\preceq\rho,\sigma\preceq\Id$, we also have
$-\Id/2\preceq S_-\preceq\Id/2$. Compression preserves order, and
$\tr\rbra{QS_+Q}=\tr\rbra{S_+Q}\leq\tr\rbra{S_+}=1$.

For the general pair, the order inequalities imply
$\ker S_+\subseteq\ker S_-$. Define $J=S_+^{-1/2}S_-S_+^{-1/2}$ on
$\supp S_+$ and set $J=0$ on $\ker S_+$. Then $J$ is Hermitian and
$-\Id\preceq J\preceq\Id$, proving \eqref{eq:factorization}.
Schatten H\"older gives
$\norm{S_-}_p\leq\norm{S_+^{1/2}}_{2p}^2\norm{J}_\infty
\leq\norm{S_+}_p$ for $1\leq p\leq\infty$.
\end{proof}

In procedure calls, density operator arguments after a semicolon denote sample access.

\begin{lemma}[{\cite[Theorem~1.2]{OW16}}]
\label{lem:tomography}
There are a universal constant $C>0$ and an algorithm
$\mathsf{Tomography}$ such that, for all integers $d,m\geq1$ and every density operator
$\omega$ on $\mathbb C^d$, $\mathsf{Tomography}\rbra{m;\omega}$ uses $m$
samples of $\omega$ and outputs a density operator $\widehat\omega$ satisfying
\begin{equation}
\E\sbra*{\norm{\widehat\omega-\omega}_2^2}\leq\frac{Cd}{m}.
\label{eq:tomography}
\end{equation}
\end{lemma}

% \begin{proof}
% The cited theorem gives the bound $\rbra{4d-3}/m$, so $C=4$ suffices.
% \end{proof}

\subsection{Absolute-value approximation}
\label{sec:absolute}

We use the following polynomial approximation to absolute value.

\begin{lemma}[Absolute-value polynomial]
\label{lem:absolute}
There is a universal constant $C>0$ such that, for every integer $K\geq2$,
there is a real even polynomial $q_K$ of degree at most $K$ satisfying
\begin{align}
q_K\rbra{0}&=0,\qquad
\sup_{\abs{x}\leq1}\abs{q_K'\rbra{x}}\leq1,\qquad
\sup_{\abs{x}\leq1}\abs[\big]{q_K\rbra{x}-\abs{x}}\leq\frac{C}{K},
\label{eq:absolute-basic}\\
\sup_{\abs{x}\leq1/2}\abs{q_K^{\rbra{j}}\rbra{x}}&\leq C^jK^{j-1},
\qquad 2\leq j\leq K,
\label{eq:absolute-derivatives}
\end{align}
and
\begin{equation}
\int_{-1}^1\abs{\sgn\rbra{x}-q_K'\rbra{x}}\,\mathrm d x\leq\frac{C}{K},
\qquad
\sup_{\abs{x}\leq1}\abs{x}\,\abs{\sgn\rbra{x}-q_K'\rbra{x}}\leq\frac{C}{K}.
\label{eq:absolute-derivative-errors}
\end{equation}
\end{lemma}

Throughout the paper, we fix the family $q_K$ constructed in
\cref{app:absolute-construction}. For $s>0$, set
$q_{K,s}\rbra{x}=s q_K\rbra{x/s}$, and for Hermitian $U$ write
$E_{K,s}\rbra{U}=\tr\rbra[\big]{\abs{U}-q_{K,s}\rbra{U}}$.

\begin{lemma}[Trace error under a low-rank perturbation]
\label{lem:trace-errors}
There is a universal constant $C>0$ such that, for every integer $K\geq2$,
$s>0$, and Hermitian matrices $U,V$ of the same size with
$\norm{U}_\infty,\norm{V}_\infty\leq s$,
\begin{equation}
\abs{E_{K,s}\rbra{U}-E_{K,s}\rbra{V}}
\leq\frac{Cs}{K}\rank\rbra{U-V}.
\label{eq:trace-rank-error}
\end{equation}
\end{lemma}

The first lemma controls the compressed bias; the second controls
the correction. Their proofs are in \cref{app:absolute}.

\subsection{The multiscale polynomial}

We approximate $\tr\rbra{\abs{Y}}$ for Hermitian $d\times d$ matrices
$X,Y$ satisfying $X\succeq0$, $\tr\rbra{X}\leq1$, and $-X\preceq Y\preceq X$.
Recall that $q_K$ from \cref{lem:absolute} has degree at most $K$ and
approximates $\abs{x}$ on $[-1,1]$.

\paragraph{Polynomial filters.}
At scale $b$, we use a filter $R_b\rbra{X}$ that is nearly the identity
on the eigenspaces of $X$ with eigenvalues at most $b$.
Let $T_m$ be the Chebyshev polynomial defined by
$T_m\rbra{\cos\theta}=\cos\rbra{m\theta}$. For $0<b<1/32$, set
\[
m=\floor*{\frac1{\sqrt{8b}}},\qquad
r_m\rbra{x}=\frac{1-T_m\rbra{1-x}}{m^2x},
\]
where $r_m$ is extended to $x=0$ by continuity, giving $r_m\rbra{0}=1$.
For $0<b\leq1$, define
\begin{equation}
R_b\rbra{x}=
\begin{cases}
1,&b\geq1/32,\\
2r_m\rbra{x}-r_m\rbra{x}^2,&b<1/32,
\end{cases}
\label{eq:filter}
\end{equation}
The numerator of $r_m$ vanishes at zero, so $r_m$ is a polynomial.

\begin{lemma}[Polynomial filter]
\label{lem:filter}
For every $0<b\leq1$, the polynomial $R_b$ has
$\deg\rbra{R_b}\leq b^{-1/2}$, $R_b\rbra{0}=1$, and
\begin{equation}
0\leq R_b\rbra{x}\leq1,\qquad xR_b\rbra{x}^2\leq2^7b
\quad\rbra{0\leq x\leq1}.
\label{eq:filter-bound}
\end{equation}
Furthermore,
\begin{equation}
-\log\rbra{R_b\rbra{x}}\leq\rbra*{\frac xb}^2
\quad\rbra{0\leq x\leq b}.
\label{eq:filter-near-zero}
\end{equation}
\end{lemma}
The proof is given in \cref{app:filter}.

\paragraph{Multiscale construction.}
For $d\geq2$ and $0<\varepsilon\leq1/32$, use
$J=\ceil{\log_2\rbra{2^9d/\varepsilon}}$ scales. Fix a sufficiently large
universal constant $C$ and set
\begin{equation}
K_\ell=\ceil*{\frac{C}{\varepsilon}\sqrt{\min\cbra{1,d2^{-\ell}}}},
\qquad 1\leq\ell\leq J.
\label{eq:scales}
\end{equation}
At scale $2^{-\ell}$, set
\[
Y_0=Y,\qquad
Y_\ell=R_{2^{-\ell}}\rbra{X}\,Y\,R_{2^{-\ell}}\rbra{X},
\qquad 1\leq\ell\leq J.
\]
Define
\begin{equation}
F_{d,\varepsilon}\rbra{X,Y}=\sum_{\ell=1}^J
2^{9-\ell}\,\tr\rbra*{
q_{K_\ell}\rbra{2^{\ell-9}Y_{\ell-1}}
-q_{K_\ell}\rbra{2^{\ell-9}Y_\ell}}.
\label{eq:multiscale}
\end{equation}
Each summand uses a rescaled absolute-value approximation for
$\tr\rbra{\abs{Y_{\ell-1}}}-\tr\rbra{\abs{Y_\ell}}$.
These target differences telescope to
$\tr\rbra{\abs{Y}}-\tr\rbra{\abs{Y_J}}$, and the choice of $J$
ensures $\norm{Y_J}_1\leq\varepsilon/4$.

\paragraph{Taylor coefficient bounds.}
For a scalar-valued polynomial $f$ of two matrix variables and a fixed expansion point
$\rbra{S_+,S_-}$, define $\Phi_j\rbra{E_0,E_1}$ to be the coefficient of $t^j$ in
\begin{equation}
 f\rbra{S_+ +tE_0,S_-+tE_1}
 =f\rbra{S_+,S_-}+\sum_{j=1}^{\deg\rbra{f}}t^j\Phi_j\rbra{E_0,E_1}.
 \label{eq:variance-taylor-expansion}
\end{equation}
Thus $\Phi_j$ is homogeneous of total degree $j$ in $E_0,E_1$.

\paragraph{Bounded representations.}
For $M_j\geq0$, we say that $\Phi_j$ has an
\emph{$M_j$-bounded representation} if there are finitely many scalars
$c_\alpha$, fixed matrices
$A_{\alpha,0},\ldots,A_{\alpha,j}\in\mathbb C^{d\times d}$, and
indices $t_{\alpha,i}\in\cbra{0,1}$ such that
\begin{equation}
 \Phi_j\rbra{E_0,E_1}
 =\sum_\alpha c_\alpha
 \tr\rbra*{A_{\alpha,0}E_{t_{\alpha,1}}A_{\alpha,1}
 \cdots E_{t_{\alpha,j}}A_{\alpha,j}}
 \label{eq:variance-taylor-representation}
\end{equation}
for all $E_0,E_1\in\mathbb C^{d\times d}$, and
\begin{equation}
 \sum_\alpha\abs{c_\alpha}
 \prod_{i=0}^j\norm{A_{\alpha,i}}_\infty\leq M_j.
 \label{eq:variance-taylor-bound}
\end{equation}
The coefficients and matrices may depend on $f,j,S_+,S_-$,
but not on $E_0,E_1$.

\begin{proposition}[Multiscale polynomial bounds]
\label{prop:multiscale}
There is a universal constant $C>0$ such that the following holds. Let
$d\geq2$ be an integer, $0<\varepsilon\leq1/32$, and $S_+,S_-$ be Hermitian
matrices on $\mathbb C^d$ satisfying $S_+\succeq0$, $\tr\rbra{S_+}\leq1$,
and $-S_+\preceq S_-\preceq S_+$. Then $F_{d,\varepsilon}$ defined in \eqref{eq:multiscale} satisfies
\begin{equation}
\deg\rbra{F_{d,\varepsilon}}\leq C\sqrt d/\varepsilon,
\qquad
\abs[\big]{F_{d,\varepsilon}\rbra{S_+,S_-}-\tr\rbra*{\abs{S_-}}}\leq\varepsilon.
\label{eq:multiscale-guarantee}
\end{equation}
For each $1\leq j\leq\deg\rbra{F_{d,\varepsilon}}$, its order-$j$
Taylor coefficient at $\rbra{S_+,S_-}$ has an $M_j$-bounded representation with
\begin{equation}
 M_1\leq\frac{C\log\rbra{d}}{\varepsilon^3},
 \qquad
 M_j\leq\frac{\rbra{Cd/\varepsilon^2}^j}{d\varepsilon\rbra{2j}!}
 \quad\rbra{2\leq j\leq\deg\rbra{F_{d,\varepsilon}}}.
 \label{eq:variance-multiscale-coefficients}
\end{equation}
\end{proposition}

The approximation and coefficient bounds are proved in
\cref{app:multiscale-approximation,app:coefficients}, respectively.

\subsection{Polynomial observables}
\label{sec:observables}

For $d\times d$ matrix variables $X,Y$, set $X_0=X$ and $X_1=Y$.
For background on trace polynomials and their degree, see
\cite{Kemp2017,Klep2022}.
Here we use a single-trace convention: a \emph{trace polynomial} is
a scalar constant plus a finite
linear combination of single-trace terms
\[
\tr\rbra{A_1X_{t_1}\cdots A_kX_{t_k}},\qquad t_i\in\cbra{0,1},
\]
where the $A_i$ are fixed known $d\times d$ matrices.
The degree of a given trace expression is the largest number $k$ of
occurrences of $X$ or $Y$ in one of its terms. Scalar constants and the
fixed matrix coefficients have degree zero.

For $\pi\in\Sym\rbra{k}$, let $U_\pi$ permute $k$ identical systems by
$U_\pi\ket{x_1,\ldots,x_k}=\ket{x_{\pi^{-1}\rbra{1}},\ldots,x_{\pi^{-1}\rbra{k}}}$,
and let $V_k=U_{\rbra{1\,2\,\cdots\,k}}^\dagger$ be the left cyclic shift
\cite[Equations~(3)--(4)]{EAO02}.
Our kernels act on bit--sample pairs in $\mathbb C^2\otimes\mathbb C^d$.
Let $W_\pi$ permute $k$ such pairs, and write
\[
\Sym_k\rbra{A}=\frac1{k!}\sum_{\pi\in\Sym\rbra{k}}W_\pi A W_\pi^\dagger.
\]
With $Z=\ketbra00-\ketbra11$ acting on the bit, associate the term
$a\tr\rbra{A_1X_{t_1}\cdots A_kX_{t_k}}$ with the kernel
\begin{equation}
a\,\Sym_k\rbra*{
\rbra*{\bigotimes_{i=1}^kZ^{t_i}}
\otimes V_k\rbra*{\bigotimes_{i=1}^k A_i}
}.
\label{eq:word-kernel}
\end{equation}
Here the tensor factors are grouped into bits and sample systems.
For $f$ real on Hermitian inputs, let $h_k$ be the Hermitian part of the
sum of its degree-$k$ kernels.

For a symmetric Hermitian kernel $h$ on $k$ pairs, its quantum
$U$-statistic \cite[Definition~6.3]{GB10} is
\begin{equation}
\mathcal U_{n,k}\rbra{h}=\binom nk^{-1}
\sum_{\substack{A\subseteq[n]\\\abs{A}=k}}h^{\rbra{A}},\qquad n\geq k,
\label{eq:u-statistic}
\end{equation}
where $h^{\rbra{A}}$ acts on the pairs in $A$ and as identity elsewhere.
For $D=\deg\rbra{f}$ and constant term $f_0$, define
\begin{equation}
G_{f,n}=f_0\Id+\sum_{k=1}^D\mathcal U_{n,k}\rbra{h_k},\qquad n\geq D.
\label{eq:observable}
\end{equation}

\begin{algorithm}[H]
\caption{$\mathsf{EstimatePolynomial}\rbra{f,n;\rho,\sigma}$}
\label{alg:polynomial}
\begin{algorithmic}[1]
\Require A known trace polynomial $f$ real on Hermitian inputs; a positive integer
$n\geq\deg\rbra{f}$; sample access to density operators $\rho,\sigma$ on $\mathbb C^d$.
\State Prepare $\tau^{\otimes n}$, where
\begin{equation}
\tau=\frac12\rbra[\big]{\ketbra00\otimes\rho+\ketbra11\otimes\sigma}
\label{eq:label-state}
\end{equation}
by choosing $\rho$ or $\sigma$ uniformly for each sample and attaching
$\ket0$ or $\ket1$, respectively.
\State Form the observable $G_{f,n}$ in \eqref{eq:observable}.
\State \Return the outcome $\widehat f_n$ of measuring $G_{f,n}$ on $\tau^{\otimes n}$.
\end{algorithmic}
\end{algorithm}

\begin{proposition}[Unbiased polynomial estimator]
\label{prop:unbiased}
Let $\rho,\sigma$ be density operators on $\mathbb C^d$, and set
$S_+=(\rho+\sigma)/2$ and $S_-=(\rho-\sigma)/2$.
Let $f\rbra{X,Y}$ be a known trace polynomial on $d\times d$ matrices,
real on Hermitian inputs, and let $n\geq\deg\rbra{f}$ be a positive integer.
Then $\mathsf{EstimatePolynomial}\rbra{f,n;\rho,\sigma}$ uses at most $n$
samples of each input and returns a real random variable $\widehat f_n$ satisfying
\begin{equation}
\E\sbra*{\widehat f_n}=f\rbra{S_+,S_-}.
\label{eq:unbiased}
\end{equation}
\end{proposition}

\begin{proof}
Expanding $\tau$ from \eqref{eq:label-state} and using the cyclic-shift identity
\[
\tr\rbra{V_k\rbra{X_1\otimes\cdots\otimes X_k}}
=\tr\rbra{X_1\cdots X_k},
\]
each word kernel has expectation equal to its trace term at $X=S_+,Y=S_-$.
Symmetrization preserves it, and taking the Hermitian part takes its real part.
Thus, for $D=\deg\rbra{f}$ and $f$ real on Hermitian inputs,
\[
\E\sbra*{\widehat f_n}
=\tr\rbra{\tau^{\otimes n}G_{f,n}}
=f_0+\sum_{k=1}^D\tr\rbra{\tau^{\otimes k}h_k}
=f\rbra{S_+,S_-}.
\]
Each pair uses one sample of either $\rho$ or $\sigma$, so at most $n$
samples of each input are needed.
\end{proof}

\section{The estimator}
\label{sec:estimator}

The estimator is given in \cref{alg:estimator}.

\begin{algorithm}[H]
\caption{$\mathsf{EstimateTraceDistance}\rbra{d,\varepsilon;\rho,\sigma}$}
\label{alg:estimator}
\begin{algorithmic}[1]
\Require An integer $d\geq2$, accuracy $0<\varepsilon\leq1/32$,
and sample access to density operators $\rho,\sigma$ on $\mathbb C^d$.
\State Fix sufficiently large universal constants $C_0,C,d_0$ as in
\cref{sec:main-proof}.
\If{$d<d_0$}
\State For each input $\omega=\rho,\sigma$, set
$\widehat\omega\gets\Call{Tomography}{\ceil{C_0d^2/\varepsilon^2};\omega}$.
\State \Return $\frac12\norm{\widehat\rho-\widehat\sigma}_1$.
\EndIf
\State Set
\begin{equation}
m_0=\ceil*{C_0d\log^2\rbra{d}},
\qquad
n=\ceil*{
\frac{Cd^2}{\varepsilon^2\log^2\rbra{d}\max\cbra{1,\varepsilon^2\log^2\rbra{d}}}
}.
\label{eq:sample-sizes}
\end{equation}
\State Prepare $m_0$ samples of $S_+=(\rho+\sigma)/2$ by choosing $\rho$ or
$\sigma$ independently and uniformly for each sample.
\State $\widehat S_+\gets\Call{Tomography}{m_0;S_+}$ using those samples.
\State Set
\begin{equation}
Q=\mathbf1_{[0,1/\rbra{2\log\rbra{d}})}\rbra{\widehat S_+},\qquad
K_{\mathrm c}=\ceil*{\frac{C_0\log\rbra{d}}{\varepsilon}}.
\label{eq:projectors}
\end{equation}
\State Define
\begin{equation}
f_{\mathrm r}\rbra{X,Y}=
\begin{cases}
\tr\rbra*{q_{K_{\mathrm r}}\rbra{Y}},\quad
K_{\mathrm r}=\ceil*{\dfrac{C_0d}{\varepsilon}},
 &\varepsilon\log\rbra{d}\leq1,\\[6pt]
F_{d,\varepsilon/8}\rbra{X,Y},&\varepsilon\log\rbra{d}>1.
\end{cases}
\label{eq:compressed}
\end{equation}
\State Define
\begin{equation}
f\rbra{X,Y}=f_{\mathrm r}\rbra{QXQ,QYQ}+\tr\rbra*{q_{K_{\mathrm c}}\rbra{Y}}
-\tr\rbra*{q_{K_{\mathrm c}}\rbra{QYQ}}.
\label{eq:main-polynomial}
\end{equation}
\State $\widehat f_n\gets\Call{EstimatePolynomial}{f,n;\rho,\sigma}$,
using samples independent of tomography.
\State \Return $\min\cbra{1,\max\cbra{0,\widehat f_n}}$.
\end{algorithmic}
\end{algorithm}

For $d<d_0$, the algorithm returns the trace distance between tomography
estimates of the two inputs. For $d\geq d_0$, it uses two independent batches:
$m_0$ samples for tomography and $n$ samples for polynomial estimation.
The estimator proceeds as follows.
\begin{enumerate}
\item \textbf{Choose the compression.}
Tomography estimates $S_+=(\rho+\sigma)/2$ and takes $Q$ to be the projector
onto the eigenspaces of $\widehat S_+$ below $1/\rbra{2\log\rbra{d}}$.
Since $\widehat S_+$ is a density operator, the discarded subspace has
dimension $\rank\rbra{\Id-Q}\leq2\log\rbra{d}$ for every outcome.
With constant probability, $\norm{QS_+Q}_\infty\leq1/\log\rbra{d}$,
as shown in \cref{sec:main-proof}.
This norm bound controls the variance of the estimate on the compressed subspace.

\item \textbf{Construct the polynomial.}
The term $f_{\mathrm r}\rbra{QS_+Q,QS_-Q}$ approximates
$\tr\rbra{\abs{QS_-Q}}$. It uses $q_{K_{\mathrm r}}$ when
$\varepsilon\log\rbra{d}\leq1$ and $F_{d,\varepsilon/8}$ otherwise.
The remaining two terms in \eqref{eq:main-polynomial} approximate
$\tr\rbra{\abs{S_-}}-\tr\rbra{\abs{QS_-Q}}$.
This correction accounts for the full change under compression, including
the off-diagonal blocks of $S_-$. Since
$\rank\rbra{S_--QS_-Q}\leq4\log\rbra{d}$,
\cref{lem:trace-errors} allows degree
$K_{\mathrm c}=O\rbra{\log\rbra{d}/\varepsilon}$ for the correction.

\item \textbf{Measure the polynomial.}
With $Q$ fixed by tomography, apply $\mathsf{EstimatePolynomial}$ to $f$
using the independent batch. The compressed term and correction form
one observable $G_{f,n}$, measured on $\tau^{\otimes n}$.
By \cref{prop:unbiased}, its outcome satisfies
\[
\E\sbra*{\widehat f_n\mid\widehat S_+}=f\rbra{S_+,S_-}.
\]
Clipping the outcome to $[0,1]$ cannot increase its error in estimating
$\dtr\rbra{\rho,\sigma}$.
\end{enumerate}
The degree bound in \cref{prop:variance} ensures that the polynomial
measurement is defined for every tomography outcome. We next bound its
approximation bias and measurement variance.

\begin{proposition}[Bias bound]
\label{prop:bias}
Let $d\geq2$, $0<\varepsilon\leq1/32$, and let $\rho,\sigma$ be density
operators on $\mathbb C^d$. Set $S_+=(\rho+\sigma)/2$ and
$S_-=(\rho-\sigma)/2$, and let $Q$ be an orthogonal projector satisfying
$\rank\rbra{\Id-Q}\leq2\log\rbra{d}$.
Define $f$ by \eqref{eq:main-polynomial} using $d,\varepsilon,Q$. Then
\begin{equation}
\abs{f\rbra{S_+,S_-}-\dtr\rbra{\rho,\sigma}}\leq\varepsilon/4.
\label{eq:bias}
\end{equation}
\end{proposition}

\begin{proof}
By \cref{lem:absolute,prop:multiscale},
\[
 \abs{f_{\mathrm r}\rbra{QS_+Q,QS_-Q}-\tr\rbra{\abs{QS_-Q}}}
 \leq
 \begin{cases}
 Cd/K_{\mathrm r},&\varepsilon\log\rbra{d}\leq1,\\
 \varepsilon/8,&\varepsilon\log\rbra{d}>1,
 \end{cases}
 \leq\varepsilon/8
\]
for sufficiently large $C_0$. Also,
\[
 \rank\rbra{S_--QS_-Q}\leq2\rank\rbra{\Id-Q}\leq4\log\rbra{d}.
\]
Hence \cref{lem:trace-errors} gives
\[
 \abs{E_{K_{\mathrm c},1}\rbra{S_-}-E_{K_{\mathrm c},1}\rbra{QS_-Q}}
 \leq\frac{C\log\rbra{d}}{K_{\mathrm c}}\leq\varepsilon/8.
\]
Using \eqref{eq:main-polynomial} and $\dtr\rbra{\rho,\sigma}=\tr\rbra{\abs{S_-}}$,
\begin{align*}
 \abs{f\rbra{S_+,S_-}-\dtr\rbra{\rho,\sigma}}
 &\leq\abs{f_{\mathrm r}\rbra{QS_+Q,QS_-Q}-\tr\rbra{\abs{QS_-Q}}}\\
 &\quad+\abs{E_{K_{\mathrm c},1}\rbra{S_-}-E_{K_{\mathrm c},1}\rbra{QS_-Q}}
 \leq\varepsilon/4.
\end{align*}
\end{proof}

The following variance bound holds for a fixed $Q$ satisfying the norm
condition obtained from tomography. Its proof is in \cref{sec:variance}.

\begin{proposition}[Variance bound]
\label{prop:variance}
There are universal constants $d_0\geq2$ and $C>0$ such that the following holds.
Let $d\geq d_0$, $0<\varepsilon\leq1/32$, and let $\rho,\sigma$ be density
operators on $\mathbb C^d$. Set $S_+=(\rho+\sigma)/2$ and
$S_-=(\rho-\sigma)/2$, let $Q$ be an orthogonal projector, and define
$f$ by \eqref{eq:main-polynomial} using $d,\varepsilon,Q$. For every integer
$n$ satisfying
\[
n\geq\frac{Cd^2}
{\varepsilon^2\log^2\rbra{d}\max\cbra{1,\varepsilon^2\log^2\rbra{d}}},
\]
we have $n\geq\deg\rbra{f}$. If also
$\norm{QS_+Q}_\infty\leq1/\log\rbra{d}$, the output $\widehat f_n$ of
$\mathsf{EstimatePolynomial}\rbra{f,n;\rho,\sigma}$ satisfies
\begin{equation}
\Var\sbra*{\widehat f_n}\leq\varepsilon^2/64.
\label{eq:variance-main}
\end{equation}
\end{proposition}

Independence of the two batches makes this bound applicable conditional on
the tomography output whenever the stated norm condition holds.

\section{Variance analysis}
\label{sec:variance}

We use quantum Fourier analysis with respect to the input product density operator
to bound the squared Fourier coefficients at each degree. Parseval's
identity then bounds the variance. Scalar derivatives control the
univariate coefficients; ordered Taylor coefficients control the
multiscale coefficients.
Observable variances in this section are taken in $\tau^{\otimes n}$,
with $\tau$ from \cref{alg:polynomial}.

\subsection{Quantum Fourier analysis and variance reduction}
\label{sec:qfa}
\label{sec:variance-tools}

Quantum Fourier analysis expands an observable in tensor products of
orthonormal single-system operators. We use the operator inner product
of \cite[Section~5]{GB10}, weighted by a density operator, so that
Parseval's identity gives the measurement variance.
Let $\tau$ be a density operator on $\mathbb C^m$. For operators $A,B$ on $r$ systems, write
\begin{equation}
 \langle A,B\rangle_{\tau,r}=\tr\rbra{\tau^{\otimes r}A^\dagger B},
 \qquad \norm{A}_{\tau,r}^2=\langle A,A\rangle_{\tau,r}.
 \label{eq:variance-weighted-norm}
\end{equation}
For full-rank $\tau$, choose an orthonormal basis
$F_0=\Id,F_1,\ldots,F_{m^2-1}$ of its complex operator space under
$\langle\cdot,\cdot\rangle_{\tau,1}$. Thus
$\tr\rbra{\tau F_a}=0$ for $a\ne0$. The product density operator makes
$F_\alpha=\bigotimes_{i=1}^rF_{\alpha_i}$, with
$\alpha\in\cbra{0,\ldots,m^2-1}^r$, an orthonormal basis on $r$ systems.
The Fourier coefficients and expansion of $A$ are
\begin{equation}
 \widehat A_\tau\rbra{\alpha}=\langle F_\alpha,A\rangle_{\tau,r},
 \qquad A=\sum_\alpha\widehat A_\tau\rbra{\alpha}F_\alpha.
 \label{eq:variance-fourier-expansion}
\end{equation}
For $\tau=\Id/m$, the inner product is the normalized Hilbert--Schmidt
inner product; choosing the Pauli basis when $m=2$ gives the usual Pauli
Fourier expansion. For general $\tau$, the basis depends on the density operator
and need not consist of Hermitian operators.

Write $\supp\rbra{\alpha}=\cbra{i:\alpha_i\ne0}$ and
$\abs{\alpha}=\abs{\supp\rbra{\alpha}}$. The degree $\abs{\alpha}$ counts
the nonidentity tensor factors in $F_\alpha$. We say that $A$ has Fourier
degree at most $D$ if $\widehat A_\tau\rbra{\alpha}=0$ whenever
$\abs{\alpha}>D$. The degree-$j$ Fourier weight and Parseval's identity are
\begin{equation}
 W_j\rbra{A}=\sum_{\abs{\alpha}=j}\abs{\widehat A_\tau\rbra{\alpha}}^2,
 \qquad \norm{A}_{\tau,r}^2=\sum_{j=0}^rW_j\rbra{A}.
 \label{eq:variance-fourier-weight}
\end{equation}
These weights are independent of the choice of the single-system basis
with $F_0=\Id$. Since $\widehat A_\tau\rbra{0}=\tr\rbra{\tau^{\otimes r}A}$,
every Hermitian $A$ satisfies
\begin{equation}
 \Var_{\tau^{\otimes r}}\sbra{A}
 =\norm{A}_{\tau,r}^2-\abs{\widehat A_\tau\rbra{0}}^2
 =\sum_{j=1}^rW_j\rbra{A}.
 \label{eq:qfa-variance}
\end{equation}
Thus bounding the variance amounts to bounding the squared Fourier
coefficients at each positive degree. The basis is used only in the
analysis and need not be known to the estimator. For singular $\tau$,
we derive the variance bounds using
$\tau_\delta=\rbra{1-\delta}\tau+\delta\Id/m$ and pass to the limit
$\delta\downarrow0$, as in \cref{lem:fourier}.

\paragraph{Effective kernels.}
For the density operator $\tau$ in \eqref{eq:label-state}, the single-pair
dimension is $m=2d$.
For a symmetric $k$-pair kernel $h_k$ and $1\leq j\leq k$, define
\begin{equation}
 \rbra{h_k}_{\tau\to j}
 =\tr_{j+1:k}\rbra*{
 h_k\rbra*{\Id^{\otimes j}\otimes\tau^{\otimes\rbra{k-j}}}
 }.
 \label{eq:variance-contraction}
\end{equation}
The following is the quantum Hoeffding decomposition of
\cite[Theorem~5.6 and Lemma~6.4]{GB10} written in Fourier coordinates;
the variance bound also appears in \cite[Lemma~4.1]{CGW26}.
\begin{lemma}[Fourier coefficients and variance]
\label{lem:fourier}
Let $\tau$ be a density operator on $\mathbb C^m$, let $n\geq1$ and
$0\leq D\leq n$ be integers, and let $h_k$ be a symmetric Hermitian kernel
on $k$ copies of $\mathbb C^m$ for $1\leq k\leq D$. For $f_0\in\mathbb R$, set
\[
G=f_0\Id+\sum_{k=1}^D\mathcal U_{n,k}\rbra{h_k},\qquad
B_j=\sum_{k=j}^D\binom{k}{j}\rbra{h_k}_{\tau\to j}.
\]
If $\tau$ has full rank, then $G$ has Fourier degree at most $D$, and for
$S=\supp\rbra{\alpha}$ with $1\leq\abs{S}=j\leq D$,
\begin{equation}
 \widehat G_\tau\rbra{\alpha}
 =\binom nj^{-1}\widehat{B_j}_\tau\rbra{\alpha_S},
 \label{eq:variance-fourier-coefficients}
\end{equation}
where $\alpha_S$ is the restriction of $\alpha$ to $S$.
Consequently, for $1\leq j\leq D$,
\begin{equation}
 W_j\rbra{G}
 =\binom nj^{-1}\sum_{a_1,\ldots,a_j\ne0}
 \abs{\widehat{B_j}_\tau\rbra{a_1,\ldots,a_j}}^2
 \leq\binom nj^{-1}\norm{B_j}_{\tau,j}^2,
 \label{eq:variance-fourier-level}
\end{equation}
and
\begin{equation}
 \Var_{\tau^{\otimes n}}\sbra*{G}
 =\sum_{j=1}^DW_j\rbra{G}
 \leq\sum_{j=1}^D\binom nj^{-1}\norm{B_j}_{\tau,j}^2.
 \label{eq:variance-fourier}
\end{equation}
The variance upper bound holds for every density operator $\tau$, including singular ones.
\end{lemma}

\begin{proof}
First assume that $\tau$ has full rank. Fix a multi-index $\alpha$ with
support $S$ of size $j\geq1$, and a subset $A\subseteq[n]$ of size $k$.
If $S\nsubseteq A$, choose $i\in S\setminus A$. The operator $h_k^{\rbra{A}}$
acts as the identity at position $i$, so the product trace defining
$\langle F_\alpha,h_k^{\rbra{A}}\rangle_{\tau,n}$ contains the factor
$\tr\rbra{\tau F_{\alpha_i}^\dagger}=0$. Thus this coefficient vanishes.
If $S\subseteq A$, symmetry of $h_k$ lets us place the positions in $S$
first. Tracing over all positions outside $A$ gives factors $\tr\rbra{\tau}=1$,
and \eqref{eq:variance-contraction} gives
\begin{align*}
 \langle F_\alpha,h_k^{\rbra{A}}\rangle_{\tau,n}
 &=\tr\rbra*{\rbra{\tau^{\otimes j}F_{\alpha_S}^\dagger
 \otimes\tau^{\otimes\rbra{k-j}}}h_k}\\
 &=\langle F_{\alpha_S},\rbra{h_k}_{\tau\to j}\rangle_{\tau,j}.
\end{align*}

For $k<j$, no $k$-subset contains $S$, so
$\langle F_\alpha,\mathcal U_{n,k}\rbra{h_k}\rangle_{\tau,n}=0$.
For $k\geq j$, exactly $\binom{n-j}{k-j}$ of the $\binom nk$ subsets contain $S$.
Therefore
\begin{align*}
 \langle F_\alpha,\mathcal U_{n,k}\rbra{h_k}\rangle_{\tau,n}
 &=\binom nk^{-1}\sum_{\substack{A\subseteq[n]\\\abs{A}=k}}
 \langle F_\alpha,h_k^{\rbra{A}}\rangle_{\tau,n}\\
 &=\frac{\binom{n-j}{k-j}}{\binom nk}
 \langle F_{\alpha_S},\rbra{h_k}_{\tau\to j}\rangle_{\tau,j}\\
 &=\frac{\binom kj}{\binom nj}
 \langle F_{\alpha_S},\rbra{h_k}_{\tau\to j}\rangle_{\tau,j}.
\end{align*}
The constant $f_0\Id$ has zero Fourier coefficients at positive degrees.
For $j\leq D$, summing over $k$ consequently yields
\begin{align*}
 \widehat G_\tau\rbra{\alpha}
 &=\binom nj^{-1}\sum_{k=j}^D\binom kj
 \langle F_{\alpha_S},\rbra{h_k}_{\tau\to j}\rangle_{\tau,j}\\
 &=\binom nj^{-1}\langle F_{\alpha_S},B_j\rangle_{\tau,j},
\end{align*}
which proves \eqref{eq:variance-fourier-coefficients}.
If $j>D$, every kernel has $k<j$, so $\widehat G_\tau\rbra{\alpha}=0$.
Hence $G$ has Fourier degree at most $D$.

For $1\leq j\leq D$, each support $S$ of size $j$ has the same coefficient
formula. Summing its squared modulus over all supports and all nonzero
single-system indices, and applying Parseval's identity
\eqref{eq:variance-fourier-weight} on $j$ systems, gives
\begin{align*}
 W_j\rbra{G}
 &=\sum_{\substack{S\subseteq[n]\\\abs{S}=j}}
 \sum_{a_1,\ldots,a_j\ne0}\binom nj^{-2}
 \abs{\widehat{B_j}_\tau\rbra{a_1,\ldots,a_j}}^2\\
 &=\binom nj^{-1}\sum_{a_1,\ldots,a_j\ne0}
 \abs{\widehat{B_j}_\tau\rbra{a_1,\ldots,a_j}}^2\\
 &\leq\binom nj^{-1}\sum_{a_1,\ldots,a_j=0}^{m^2-1}
 \abs{\widehat{B_j}_\tau\rbra{a_1,\ldots,a_j}}^2
 =\binom nj^{-1}\norm{B_j}_{\tau,j}^2.
\end{align*}
This proves \eqref{eq:variance-fourier-level}.
Applying \eqref{eq:qfa-variance} to $G$ gives
\[
 \Var_{\tau^{\otimes n}}\sbra*{G}
 =\sum_{j=1}^nW_j\rbra{G}
 =\sum_{j=1}^DW_j\rbra{G}
 \leq\sum_{j=1}^D\binom nj^{-1}\norm{B_j}_{\tau,j}^2.
\]

For a possibly singular $\tau$, set
$\tau_\delta=\rbra{1-\delta}\tau+\delta\Id/m$, where $0<\delta<1$.
Keep $G$ and the kernels $h_k$ fixed, and define
\[
 B_{j,\delta}=\sum_{k=j}^D\binom kj\rbra{h_k}_{\tau_\delta\to j}.
\]
Since $\tau_\delta$ has full rank, the bound just proved gives
\[
 \Var_{\tau_\delta^{\otimes n}}\sbra*{G}
 \leq\sum_{j=1}^D\binom nj^{-1}
 \tr\rbra{\tau_\delta^{\otimes j}B_{j,\delta}^\dagger B_{j,\delta}}.
\]
The partial-trace formula \eqref{eq:variance-contraction} implies
$B_{j,\delta}\to B_j$ as $\delta\downarrow0$. Continuity of the finite-dimensional
traces gives
\begin{align*}
 \Var_{\tau_\delta^{\otimes n}}\sbra*{G}
 &\longrightarrow\Var_{\tau^{\otimes n}}\sbra*{G},\\
 \tr\rbra{\tau_\delta^{\otimes j}B_{j,\delta}^\dagger B_{j,\delta}}
 &\longrightarrow\norm{B_j}_{\tau,j}^2.
\end{align*}
Taking the limit proves the inequality in \eqref{eq:variance-fourier}
for every $\tau$.
\end{proof}

Let $\mathcal C_j$ be the set of $j$-cycles in $\Sym\rbra{j}$, with
$\mathcal C_1$ containing the identity. The Hermitian cycle average is
\begin{equation}
 C_j=\frac1{\rbra{j-1}!}\sum_{c\in\mathcal C_j}U_c.
 \label{eq:variance-cycle-average}
\end{equation}
Here $U_c$ permutes the $j$ quantum registers.

\begin{lemma}[Cycle moment; {\cite[proof of Lemma~4.3]{CGW26}}]
\label{lem:cycles}
Let $R\succeq0$ satisfy $\tr\rbra{R}\leq1$ and
$\norm{R}_\infty\leq\beta$, where $\beta>0$. For every integer $j\geq1$,
the cycle average $C_j$ in \eqref{eq:variance-cycle-average} satisfies
\begin{equation}
 \tr\rbra{R^{\otimes j}C_j^2}
 \leq\min\cbra*{1,
 \frac{\prod_{\ell=0}^{j-1}\rbra{1+\ell\beta}}{\rbra{j-1}!}}.
 \label{eq:variance-cycle-bound}
\end{equation}
\end{lemma}

\begin{proof}
The argument in \cite[proof of Lemma~4.3]{CGW26} gives
\[
 \tr\rbra{T^{\otimes j}C_j^2}
 \leq\frac{\prod_{\ell=0}^{j-1}\rbra{\tr\rbra{T}+\ell}}{\rbra{j-1}!},
 \qquad 0\preceq T\preceq\Id.
\]
Taking $T=R/\beta$ and multiplying by $\beta^j$ gives
\[
 \tr\rbra{R^{\otimes j}C_j^2}
 \leq\frac{\prod_{\ell=0}^{j-1}\rbra{\tr\rbra{R}+\ell\beta}}{\rbra{j-1}!}
 \leq\frac{\prod_{\ell=0}^{j-1}\rbra{1+\ell\beta}}{\rbra{j-1}!}.
\]
Since $C_j$ is an average of unitaries, $\norm{C_j}_\infty\leq1$, and
\[
 \tr\rbra{R^{\otimes j}C_j^2}
 \leq\norm{C_j}_\infty^2\rbra{\tr\rbra{R}}^j\leq1.
\]
Together these inequalities give \eqref{eq:variance-cycle-bound}.
\end{proof}

\subsection{Taylor coefficients and contracted kernels}

\begin{proposition}[Variance from Taylor coefficients]
\label{prop:taylor-variance}
Let $\rho,\sigma$ be density operators on $\mathbb C^d$, and set
$S_+=(\rho+\sigma)/2$ and $S_-=(\rho-\sigma)/2$.
Let $Q$ be an orthogonal projector, and let $\beta>0$ satisfy
$\norm{QS_+Q}_\infty\leq\beta$.
Let $f\rbra{X,Y}$ be a trace polynomial on $d\times d$ matrices,
real on Hermitian inputs, and set $D=\deg\rbra{f}$.
Suppose that, at $\rbra{QS_+Q,QS_-Q}$, the order-$j$ Taylor coefficient
of $f$ has an $M_j$-bounded representation for each $1\leq j\leq D$.
Then, for every positive integer $n\geq D$, the observable $G_{f,Q,n}$
associated with $\rbra{X,Y}\mapsto f\rbra{QXQ,QYQ}$ satisfies
\begin{equation}
 \Var\sbra*{G_{f,Q,n}}
 \leq\sum_{j=1}^D\binom nj^{-1}M_j^2
 \frac{\prod_{\ell=0}^{j-1}\rbra{1+\ell\beta}}{\rbra{j-1}!}.
 \label{eq:variance-from-taylor}
\end{equation}
\end{proposition}

\begin{proof}
First take $Q=\Id$ and define
$\mathsf X\rbra{E_0,E_1}=\tfrac12\diag\rbra{E_0+E_1,E_0-E_1}$,
so that $\mathsf X\rbra{S_+,S_-}=\tau$.
Let $f_k$ be the homogeneous degree-$k$ part of $f$, with its original
$k$-pair kernel $h_k$. The kernel identity and its Taylor expansion give
\begin{align}
 \tr\rbra*{h_k\mathsf X\rbra{E_0,E_1}^{\otimes k}}
 &=f_k\rbra{E_0,E_1},
 \label{eq:variance-homogeneous-identity}\\
 \tr\rbra*{B_j\mathsf X\rbra{E_0,E_1}^{\otimes j}}
 &=\Phi_j\rbra{E_0,E_1}.
 \label{eq:variance-polarization-identity}
\end{align}
Each $f_k$ is real on Hermitian inputs, as follows by scaling both
inputs. Thus \eqref{eq:variance-homogeneous-identity} follows from
\eqref{eq:word-kernel} and polynomial continuation in Hermitian basis coordinates.
Since $h_k$ is symmetric, expanding at $\rbra{S_+,S_-}$ gives
\begin{align*}
 f_k\rbra{S_++tE_0,S_-+tE_1}-f_k\rbra{S_+,S_-}
 =\sum_{j=1}^k\binom kj t^j
 \tr\rbra*{\rbra{h_k}_{\tau\to j}\mathsf X\rbra{E_0,E_1}^{\otimes j}}.
\end{align*}
Summing over $k$ and comparing coefficients of $t^j$ yields
\eqref{eq:variance-polarization-identity}, with
$B_j=\sum_{k=j}^D\binom kj\rbra{h_k}_{\tau\to j}$.
Polarization on the block-diagonal algebra uniquely identifies the
symmetric kernel: the coefficient of $t_1\cdots t_j$ at
$\sum_i t_i\xi_i$ is $j!\tr\rbra{B_j\rbra{\xi_1\otimes\cdots\otimes\xi_j}}$.
Thus applying \eqref{eq:word-kernel} to any representation
\eqref{eq:variance-taylor-representation} gives the actual $B_j$.

Symmetrization over the $j!$ permutations of the whole pairs has $j$
cyclic origins for each $c\in\mathcal C_j$. Averaging these origins
and combining the two boundary factors of each trace yields
\begin{equation}
 B_j=\frac1{\rbra{j-1}!}\sum_{c\in\mathcal C_j}K_c,
 \label{eq:variance-kernel-cycle-form}
\end{equation}
where each $K_c$ is a sum of terms
\[
 a\rbra*{\bigotimes_{i=1}^jZ^{t_i}}
 \otimes\rbra*{\bigotimes_{i=1}^jA_i}U_c,
 \qquad \sum\abs a\prod_{i=1}^j\norm{A_i}_\infty\leq M_j.
\]
In $\tr\rbra{\tau^{\otimes j}K_c^\dagger K_{c'}}$, moving a permutation
past known tensor factors only permutes those factors, and tracing a
label leaves $S_+$ or $S_-$.
Along a cycle of the relative permutation $\pi=c^{-1}c'$ of length $k$, Schatten
H\"older and \eqref{eq:factorization} give
\begin{equation}
 \abs{\tr\rbra{X_1A_1\cdots X_kA_k}}
 \leq\prod_{i=1}^k\norm{X_i}_k\prod_{i=1}^k\norm{A_i}_\infty
 \leq\tr\rbra{S_+^k}\prod_{i=1}^k\norm{A_i}_\infty,
 \quad X_i\in\cbra{S_+,S_-}.
 \label{eq:variance-cycle-holder}
\end{equation}
Each known matrix contributes its operator norm exactly once. Hence
\[
 \abs{\tr\rbra{\tau^{\otimes j}K_c^\dagger K_{c'}}}
 \leq M_j^2\prod_{\gamma\in\operatorname{cycles}\rbra{\pi}}\tr\rbra{S_+^{\abs{\gamma}}}
 =M_j^2\tr\rbra{S_+^{\otimes j}U_\pi}.
\]
Using \eqref{eq:variance-kernel-cycle-form},
\begin{equation}
\begin{aligned}
 \norm{B_j}_{\tau,j}^2
 &=\frac1{\rbra{\rbra{j-1}!}^2}
 \sum_{c,c'\in\mathcal C_j}\tr\rbra{\tau^{\otimes j}K_c^\dagger K_{c'}}\\
 &\leq\frac1{\rbra{\rbra{j-1}!}^2}
 \sum_{c,c'\in\mathcal C_j}\abs{\tr\rbra{\tau^{\otimes j}K_c^\dagger K_{c'}}}\\
 &\leq\frac{M_j^2}{\rbra{\rbra{j-1}!}^2}
 \sum_{c,c'\in\mathcal C_j}\tr\rbra{S_+^{\otimes j}U_{c^{-1}c'}}
 =M_j^2\tr\rbra{S_+^{\otimes j}C_j^2}.
\end{aligned}
 \label{eq:variance-cycle-comparison}
\end{equation}
Combining \eqref{eq:variance-fourier} with \eqref{eq:variance-cycle-comparison}
and \cref{lem:cycles} gives
\begin{align*}
 \Var\sbra*{G_{f,n}}
 &\leq\sum_{j=1}^D\binom nj^{-1}M_j^2
 \tr\rbra{S_+^{\otimes j}C_j^2}\\
 &\leq\sum_{j=1}^D\binom nj^{-1}M_j^2
 \frac{\prod_{\ell=0}^{j-1}\rbra{1+\ell\beta}}{\rbra{j-1}!}.
\end{align*}

For general $Q$, every quantum position has $Q$ on both sides.
Contractions and uncentered squared norms therefore use $QS_+Q,QS_-Q$,
so \eqref{eq:variance-cycle-comparison} holds with $QS_+Q$ in place of
$S_+$. Since $\tr\rbra{QS_+Q}\leq1$ and $-QS_+Q\preceq QS_-Q\preceq QS_+Q$, the
same bounds apply. The Fourier expansion uses the original normalized
density operator $\tau$; no division by $\tr\rbra{QS_+Q}$ is introduced.
\end{proof}

\subsection{Univariate approximation}

\begin{lemma}[Univariate variance]
\label{lem:univariate}
There is a universal constant $C>0$ such that the following holds.
Let $\rho,\sigma$ be density operators on $\mathbb C^d$, set
$S_+=(\rho+\sigma)/2$, and let $Q$ be an orthogonal projector.
Let $K,n$ be integers with $K\geq2$ and $n\geq2K$, and let $G_{K,Q,n}$
be the observable \eqref{eq:observable} for
$\rbra{X,Y}\mapsto\tr\rbra{q_K\rbra{QYQ}}$, with variance in
$\tau^{\otimes n}$ for $\tau$ in \eqref{eq:label-state}. Then
\begin{equation}
 \Var\sbra*{G_{K,Q,n}}
 \leq\frac{1}{K^2}
 \rbra*{\exp\rbra*{\frac{CK^2}{n}}-1}.
 \label{eq:variance-univariate-crude}
\end{equation}
Moreover, if $\norm{QS_+Q}_\infty\leq\beta$ for some $\beta>0$, then
\begin{equation}
 \Var\sbra*{G_{K,Q,n}}
 \leq\frac{1}{n}+\frac{1}{K^2}
 \rbra*{
 \exp\rbra*{\frac{CK}{\sqrt n}}
 +\exp\rbra*{\frac{C\beta K^2}{n}}
 }.
 \label{eq:variance-univariate}
\end{equation}
\end{lemma}

\begin{proof}
Write $q_K\rbra{x}=\sum_{k=1}^K a_kx^k$.
Symmetrizing the cyclic shift in \eqref{eq:word-kernel} shows that
the degree-$k$ kernel for $\tr\rbra{q_K\rbra{QYQ}}$ is
\[
 h_k=a_kZ^{\otimes k}\otimes Q^{\otimes k}C_kQ^{\otimes k}.
\]
For a contracted pair, tracing over the bit gives
\[
 \tr_{\mathbb C^2}\rbra*{\rbra{Z\otimes Q}\tau\rbra{\Id_2\otimes Q}}
 =\frac12 Q\rbra{\rho-\sigma}Q=QS_-Q.
\]
For a density operator $\omega$ and $1\leq j\leq k$,
\cite[proof of Lemma~4.2]{CGW26} establishes the contraction identity
\[
 \binom kj\tr_{j+1:k}\rbra*{
 C_k\rbra{\Id^{\otimes j}\otimes\omega^{\otimes\rbra{k-j}}}}
 =\frac kj\rbra*{
 \sum_{\substack{r_1+\cdots+r_j=k-j\\r_i\geq0}}
 \bigotimes_{i=1}^j\omega^{r_i}}C_j.
\]
Both sides are homogeneous polynomials of degree $k-j$ in $\omega$.
Rescaling extends the identity to positive definite matrices, and
polynomial continuation on the Hermitian matrix space extends it to
every Hermitian matrix. Applying it to $QS_-Q$ after the bit trace above gives,
for $1\leq j\leq k\leq K$,
\[
 \binom kj\rbra{h_k}_{\tau\to j}
 =\frac{ka_k}{j}Z^{\otimes j}\otimes
 Q^{\otimes j}\rbra*{
 \sum_{\substack{r_1+\cdots+r_j=k-j\\r_i\geq0}}
 \bigotimes_{i=1}^j\rbra{QS_-Q}^{r_i}}C_jQ^{\otimes j}.
\]
Let $\rbra{QS_-Q}_i$ act on position $i$, and let $t=\rbra{t_1,\ldots,t_j}$ be uniform on
$\cbra{t_i\geq0:\sum_i t_i=1}$, with $t_1=1$ for $j=1$.
Since these $j$ matrix copies commute, the simplex-moment identity
\cite[Eqs.~(20) and~(23)]{CGW26} gives
\[
 \binom kj\E_t\sbra*{\rbra*{\sum_{i=1}^j t_i\rbra{QS_-Q}_i}^{k-j}}
 =\frac kj\sum_{\substack{r_1+\cdots+r_j=k-j\\r_i\geq0}}
 \bigotimes_{i=1}^j\rbra{QS_-Q}^{r_i}.
\]
Using
$q_K^{\rbra{j}}\rbra{x}/j!=\sum_{k=j}^K a_k\binom kj x^{k-j}$
and $B_j=\sum_{k=j}^K\binom kj\rbra{h_k}_{\tau\to j}$ from
\cref{lem:fourier}, we obtain
\begin{equation}
B_j=Z^{\otimes j}\otimes Q^{\otimes j}D_jC_jQ^{\otimes j},
\qquad
D_j=\frac1{j!}\E_t\sbra*{q_K^{\rbra{j}}\rbra*{\sum_{i=1}^j t_i\rbra{QS_-Q}_i}},
\label{eq:variance-univariate-kernel}
\end{equation}
Symmetry of the simplex law gives $D_jC_j=C_jD_j$. Since
$\norm{QS_-Q}_\infty\leq1/2$,
\[
 0\preceq D_j^2C_j^2
 \preceq
 \rbra*{\frac{\sup_{\abs{x}\leq1/2}\abs{q_K^{\rbra{j}}\rbra{x}}}{j!}}^2 C_j^2.
\]
Taking the trace against $\rbra{QS_+Q}^{\otimes j}$ and using $Z^2=\Id$ gives
\begin{align*}
 \norm{B_j}_{\tau,j}^2
 &=\tr\rbra{\rbra{QS_+Q}^{\otimes j}D_j^2C_j^2}\\
 &\leq\rbra*{\frac{\sup_{\abs{x}\leq1/2}\abs{q_K^{\rbra{j}}\rbra{x}}}{j!}}^2
 \tr\rbra{\rbra{QS_+Q}^{\otimes j}C_j^2}.
\end{align*}
Substituting into \eqref{eq:variance-fourier} yields
\begin{equation}
 \Var\sbra*{G_{K,Q,n}}
 \leq\sum_{j=1}^K\binom{n}{j}^{-1}
 \rbra*{\frac{\sup_{\abs{x}\leq1/2}\abs{q_K^{\rbra{j}}\rbra{x}}}{j!}}^2
 \tr\rbra{\rbra{QS_+Q}^{\otimes j}C_j^2}.
 \label{eq:variance-univariate-series}
\end{equation}
This does not require $QS_+Q$ and $QS_-Q$ to commute.

By \cref{lem:absolute},
$\sup_{\abs{x}\leq1/2}\abs{q_K^{\rbra{j}}\rbra{x}}
\leq C^jK^{j-1}$ for $1\leq j\leq K$.
Using $\tr\rbra{\rbra{QS_+Q}^{\otimes j}C_j^2}\leq1$ and
$\binom{n}{j}^{-1}\leq j!\rbra{2/n}^j$ in
\eqref{eq:variance-univariate-series} gives
\[
 \Var\sbra*{G_{K,Q,n}}
 \leq\frac1{K^2}\sum_{j\geq1}\frac{\rbra{CK^2/n}^j}{j!},
\]
which proves \eqref{eq:variance-univariate-crude}.

Now assume $\norm{QS_+Q}_\infty\leq\beta$.
Since $C_1=\Id$ and $\sup_{[-1/2,1/2]}\abs{q_K'}\leq1$,
\[
 \binom n1^{-1}
 \rbra*{\sup_{\abs{x}\leq1/2}\abs{q_K'\rbra{x}}}^2
 \tr\rbra{QS_+Q\,C_1^2}
 \leq\frac{\tr\rbra{QS_+Q}}{n}\leq\frac1n.
\]
For $2\leq j\leq K$, the derivative bound and \cref{lem:cycles} give
\begin{align*}
 &\binom nj^{-1}
 \rbra*{\frac{\sup_{\abs{x}\leq1/2}\abs{q_K^{\rbra{j}}\rbra{x}}}{j!}}^2
 \tr\rbra{\rbra{QS_+Q}^{\otimes j}C_j^2}\\
 &\qquad\leq\frac1{K^2}
 \frac{\rbra{CK^2/n}^j\prod_{\ell=0}^{j-1}\rbra{1+\ell\beta}}
 {j!\rbra{j-1}!}.
\end{align*}
Thus \eqref{eq:variance-univariate-series} implies
\begin{equation}
 \Var\sbra*{G_{K,Q,n}}\leq\frac1n+\frac{1}{K^2}
 \sum_{j\geq2}\frac{\rbra{CK^2/n}^j\prod_{\ell=0}^{j-1}\rbra{1+\ell\beta}}{j!\rbra{j-1}!}.
 \label{eq:variance-univariate-tail}
\end{equation}
For $x\geq0$ and every positive integer $a$,
\begin{equation}
\sum_{j\geq0}\frac{x^j}{\rbra{j!}^a}\leq\exp\rbra{a x^{1/a}}.
\label{eq:variance-exponential-series}
\end{equation}
Indeed, expand $\exp\rbra{x^{1/a}}^a$ and retain only terms with equal
indices. Also,
\[
 \prod_{\ell=0}^{j-1}\rbra{1+\ell\beta}
 \leq2^j\rbra[\big]{1+\rbra{j\beta}^j},\qquad
 j^j\leq e^j j!,\qquad j\leq2^j.
\]
These inequalities and \eqref{eq:variance-exponential-series}, with $a=2,1$, give
\begin{align*}
 &\sum_{j\geq2}
 \frac{\rbra{CK^2/n}^j\prod_{\ell=0}^{j-1}\rbra{1+\ell\beta}}
 {j!\rbra{j-1}!}\\
 &\qquad\leq\sum_{j\geq2}\frac{\rbra{CK^2/n}^j}{\rbra{j!}^2}
 +\sum_{j\geq2}\frac{\rbra{C\beta K^2/n}^j}{j!}\\
 &\qquad\leq\exp\rbra{CK/\sqrt n}+\exp\rbra{C\beta K^2/n}.
\end{align*}
Combining this with \eqref{eq:variance-univariate-tail} proves
\eqref{eq:variance-univariate}.
\end{proof}

\subsection{Multiscale approximation}

The coefficient estimates from \cref{prop:multiscale} yield the following
variance bound.

\begin{proposition}[Multiscale variance]
\label{prop:coefficients}
There is a universal constant $C>0$ such that the following holds.
Let $d\geq2$ be an integer, let $0<\varepsilon\leq1/32$, and let
$F_{d,\varepsilon}$ be the trace polynomial defined in \eqref{eq:multiscale}.
Set $D=\deg\rbra{F_{d,\varepsilon}}$.

Let $\rho,\sigma$ be density operators on $\mathbb C^d$, and set
$S_+=(\rho+\sigma)/2$ and $S_-=(\rho-\sigma)/2$.
Let $Q$ be an orthogonal projector, and let $\beta>0$ satisfy
$\norm{QS_+Q}_\infty\leq\beta$.
For every integer $n\geq2D$, the compressed observable satisfies
\begin{equation}
 \Var\sbra*{G_{F_{d,\varepsilon},Q,n}}
 \leq\frac{C\log^2\rbra{d}}{\varepsilon^6n}
 +\frac{C}{d^2\varepsilon^2}
 \rbra*{\exp\rbra*{\frac{C\sqrt d}{\varepsilon n^{1/4}}}
 +\exp\rbra*{C\rbra*{\frac{\beta d^2}{\varepsilon^4n}}^{1/3}}}.
 \label{eq:variance-multiscale}
\end{equation}
\end{proposition}

\begin{proof}
By \cref{prop:multiscale}, the Taylor coefficients at
$\rbra{QS_+Q,QS_-Q}$ have bounded representations satisfying
\eqref{eq:variance-multiscale-coefficients}.
In particular,
\[
 \frac{M_1^2}{n}\leq\frac{C\log^2\rbra{d}}{\varepsilon^6n}.
\]
For $2\leq j\leq D$, the binomial bound gives
\begin{align*}
 \binom nj^{-1}\frac{M_j^2}{\rbra{j-1}!}
 &\leq\frac{j!\rbra{2/n}^j}{\rbra{j-1}!}
 \frac{\rbra{Cd/\varepsilon^2}^{2j}}{d^2\varepsilon^2\rbra{\rbra{2j}!}^2}\\
 &\leq\frac{j\rbra{Cd^2/\rbra{\varepsilon^4n}}^j}
 {d^2\varepsilon^2\rbra{\rbra{2j}!}^2}.
\end{align*}
Substituting these bounds into \eqref{eq:variance-from-taylor} yields
\begin{equation}
 \Var\sbra*{G_{F_{d,\varepsilon},Q,n}}
 \leq\frac{C\log^2\rbra{d}}{\varepsilon^6n}
 +\frac{C}{d^2\varepsilon^2}
 \sum_{j\geq2}
 \frac{j\rbra{Cd^2/\rbra{\varepsilon^4n}}^j
 \prod_{\ell=0}^{j-1}\rbra{1+\ell\beta}}{\rbra{\rbra{2j}!}^2}.
 \label{eq:variance-multiscale-series}
\end{equation}
Using $\rbra{2j}!\geq2^j\rbra{j!}^2$,
$\prod_{\ell=0}^{j-1}\rbra{1+\ell\beta}\leq2^j\rbra{1+\rbra{j\beta}^j}$,
$j\leq2^j$, and $j^j\leq e^j j!$, we obtain
\begin{align*}
 &\sum_{j\geq2}
 \frac{j\rbra{Cd^2/\rbra{\varepsilon^4n}}^j
 \prod_{\ell=0}^{j-1}\rbra{1+\ell\beta}}{\rbra{\rbra{2j}!}^2}\\
 &\qquad\leq\sum_{j\geq2}\frac{\rbra{Cd^2/\rbra{\varepsilon^4n}}^j}{\rbra{j!}^4}
 +\sum_{j\geq2}\frac{\rbra{C\beta d^2/\rbra{\varepsilon^4n}}^j}{\rbra{j!}^3}.
\end{align*}
Apply \eqref{eq:variance-exponential-series} with $a=4,3$ to obtain
\eqref{eq:variance-multiscale}.
\end{proof}

\subsection{Proof of Proposition~\ref{prop:variance}}

\begin{proof}[Proof of Proposition~\ref{prop:variance}]
Fix $C_0$ sufficiently large for \cref{prop:bias}, and choose the constant
in the sample lower bound sufficiently large.
Let $G_{\mathrm r}$ be the observable for the compressed term.
By the degree choices and \cref{prop:multiscale},
$K_{\mathrm r}=O\rbra{d/\varepsilon}$,
$K_{\mathrm c}=O\rbra{\log\rbra{d}/\varepsilon}$, and
$\deg\rbra{F_{d,\varepsilon/8}}=O\rbra{\sqrt d/\varepsilon}$.
For sufficiently large $d_0$, the sample bound therefore ensures
\[
 n\geq
 \begin{cases}
 2\max\cbra{K_{\mathrm c},K_{\mathrm r}},&\varepsilon\log\rbra{d}\leq1,\\
 2\max\cbra{K_{\mathrm c},\deg\rbra{F_{d,\varepsilon/8}}},
 &\varepsilon\log\rbra{d}>1,
 \end{cases}
 \qquad \frac{CK_{\mathrm c}^2}{n}\leq1.
\]
In particular, $n\geq2\deg\rbra{f}$.

Apply \eqref{eq:variance-univariate} in the first regime and
\eqref{eq:variance-multiscale} with accuracy $\varepsilon/8$ in the second,
using $\beta=1/\log\rbra{d}$ in both cases.
For a sufficiently large constant in the sample bound, the degree choices give
\[
 \max\cbra*{\frac{CK_{\mathrm r}}{\sqrt n},
 \frac{CK_{\mathrm r}^2}{n\log\rbra{d}}}
 \leq\log\rbra{d}
 \qquad\rbra{\varepsilon\log\rbra{d}\leq1}
\]
and
\[
 \max\cbra*{\frac{C\sqrt d}{\varepsilon n^{1/4}},
 C\rbra*{\frac{d^2}{\varepsilon^4n\log\rbra{d}}}^{1/3}}
 \leq\log\rbra{d}
 \qquad\rbra{\varepsilon\log\rbra{d}>1}.
\]
Thus each exponential in the applicable variance bound is at most $d$.
Substituting the degree and sample bounds, and using
$\varepsilon^{-1}<\log\rbra{d}$ in the second regime, gives
\begin{equation}
 \frac{\Var\sbra*{G_{\mathrm r}}}{\varepsilon^2}
 \leq C
 \begin{cases}
 \dfrac{\log^2\rbra{d}}{d^2}+\dfrac{1}{d},
    &\varepsilon\log\rbra{d}\leq1,\\[6pt]
 \dfrac{\log^{10}\rbra{d}}{d^2}
       +\dfrac{\log^4\rbra{d}}{d},
    &\varepsilon\log\rbra{d}>1,
 \end{cases}
 \quad\leq\frac1{256},
 \label{eq:variance-compressed}
\end{equation}
Taking $d_0$ sufficiently large ensures
$\Var\sbra*{G_{\mathrm r}}\leq\varepsilon^2/256$.
For $P\in\cbra{\Id,Q}$, apply \eqref{eq:variance-univariate-crude} and
$\exp\rbra{t}-1\leq2t$ for $0\leq t\leq1$ to obtain
\[
 \Var\sbra*{G_{K_{\mathrm c},P,n}}
 \leq\frac1{K_{\mathrm c}^2}
 \rbra*{\exp\rbra{CK_{\mathrm c}^2/n}-1}
 \leq\frac Cn\leq\frac{\varepsilon^2}{1024},
\]
again for sufficiently large $d_0$.
Since all observables act on the same fresh samples and
$G_{f,n}=G_{\mathrm r}+G_{K_{\mathrm c},\Id,n}-G_{K_{\mathrm c},Q,n}$,
the triangle inequality for centered operators yields
\[
 \sqrt{\Var\sbra*{G_{f,n}}}
 \leq\sqrt{\Var\sbra*{G_{\mathrm r}}}
      +\sqrt{\Var\sbra*{G_{K_{\mathrm c},\Id,n}}}
      +\sqrt{\Var\sbra*{G_{K_{\mathrm c},Q,n}}}
 \leq\frac{\varepsilon}{16}+\frac{2\varepsilon}{32}
 =\frac{\varepsilon}{8}.
\]
\end{proof}

\section{Proof of the main theorem}
\label{sec:main-proof}

\begin{proof}[Proof of \cref{thm:main}]
Assume $0<\varepsilon\leq1/32$.
Choose $C_0$ sufficiently large for \cref{prop:bias} and the tomography
estimates below, then fix $C,d_0$ as in \cref{prop:variance}.

Suppose $d\geq d_0$. By \cref{lem:tomography},
Markov's inequality, and
$\norm{X}_\infty\leq\norm{X}_2$,
\begin{equation}
\Pr\sbra*{\norm{\widehat S_+-S_+}_\infty>\frac1{4\log\rbra{d}}}
\leq\frac{Cd\log^2\rbra{d}}{m_0}\leq\frac14.
\label{eq:tomography-event}
\end{equation}
Since $\widehat S_+$ is a density operator,
$\rank\rbra{\Id-Q}\leq2\log\rbra{d}$ for every output.
Outside the event in \eqref{eq:tomography-event}, the definition of $Q$
and the triangle inequality give
$\norm{QS_+Q}_\infty\leq3/\rbra{4\log\rbra{d}}\leq1/\log\rbra{d}$.

Conditional on such an output, \cref{prop:unbiased} and
\cref{prop:bias,prop:variance} give bias at most $\varepsilon/4$ and,
by Chebyshev's inequality,
$\Pr\sbra{\abs{\widehat f_n-f\rbra{S_+,S_-}}>\varepsilon/2\mid\widehat S_+}\leq1/16$.
The error is therefore at most $3\varepsilon/4$ except with probability at most
$1/4+1/16<1/3$. Clipping to $[0,1]$ cannot increase the distance to
$\dtr\rbra{\rho,\sigma}\in[0,1]$.

The estimator uses at most $m_0+n$ samples of each density operator for any labels.
Since $m_0/n=O\rbra{\log^6\rbra{d}/d}$ uniformly for $0<\varepsilon\leq1/32$,
the total has the order claimed in \cref{thm:main}.

For $2\leq d<d_0$, apply $\mathsf{Tomography}$ to $\ceil{C_0d^2/\varepsilon^2}$ samples
of each input. By \cref{lem:tomography},
$\norm{X}_1^2\leq d\norm{X}_2^2$,
and the reverse triangle inequality,
$\E\sbra{\abs{\dtr\rbra{\widehat\rho,\widehat\sigma}-\dtr\rbra{\rho,\sigma}}^2}
\leq\varepsilon^2/4$.
Markov's inequality bounds the failure probability by $1/4$.
For bounded $d$, the cost $O\rbra{d^2/\varepsilon^2}$ also has the order claimed in \cref{thm:main}.

\end{proof}

\section*{AI Use Disclosure}

ChatGPT-6 Astra was used to assist the proof of \cref{thm:main}. 
The author verified the correctness and originality of all content, including references.

\phantomsection
\addcontentsline{toc}{section}{References}
\bibliographystyle{alphaurl}
\bibliography{main}

\appendix

\section{Proof of the absolute-value polynomial bounds}
\label[appendix]{app:absolute}

\subsection{Construction and scalar approximation}
\label[appendix]{app:absolute-construction}

For integers $M\geq2$, use the normalized Jackson kernel of \cite[Definition~9.5]{Sha05}:
\begin{equation}
J_M\rbra{t}=\frac{3}{2\pi M\rbra{2M^2+1}}
\rbra*{\frac{\sin\rbra{Mt/2}}{\sin\rbra{t/2}}}^4,
\qquad \int_{-\pi}^{\pi}J_M\rbra{t}\,\mathrm d t=1.
\label{eq:jackson}
\end{equation}
It is even, nonnegative, and has trigonometric degree $2\rbra{M-1}$.
For $\abs{t}\leq\pi$ and $u>0$, the inequalities
$\abs{\sin\rbra{Mt/2}}\leq M\abs{\sin\rbra{t/2}}$ and
$\abs{\sin\rbra{t/2}}\geq\abs{t}/\pi$ give
\begin{equation}\label{eq:app-jackson-tail}
 J_M\rbra{t}\leq\min\cbra*{M,\frac{3\cdot2^3}{M^3\abs{t}^4}},\qquad
 \int_{u\leq\abs{t}\leq\pi}J_M\rbra{t}\,\mathrm d t
 \leq\min\cbra{1,2^4\rbra{Mu}^{-3}}.
\end{equation}
Splitting the first moment at $2/M$ gives the explicit version of
\cite[Lemma~9.6]{Sha05}:
\begin{equation}
\int_{-\pi}^{\pi}\abs{t}J_M\rbra{t}\,\mathrm d t
\leq\frac{10}{M}\leq\frac{2^4}{M}.
\label{eq:jackson-bounds}
\end{equation}

For an integer $K\geq2$, put $M=\floor{K/2}+1\geq K/2$ and define
\begin{equation}
h_M\rbra{\theta}=\int_{-\pi}^{\pi}J_M\rbra{t}\sgn\rbra{\cos\rbra{\theta-t}}\,\mathrm d t,
\qquad q_K\rbra{x}=\int_0^x h_M\rbra{\arccos u}\,\mathrm d u.
\label{eq:absolute-polynomial}
\end{equation}
The expression on $[-1,1]$ defines an algebraic polynomial, which we
extend to all $x$.

\begin{proof}[Proof of \cref{lem:absolute}]
The kernel has trigonometric degree at most $K$. Its convolution $h_M$
with $\sgn\rbra{\cos\theta}$ is even,
satisfies $h_M\rbra{\theta+\pi}=-h_M\rbra{\theta}$, and has absolute value at most
one. Hence $h_M\rbra{\theta}=g\rbra{\cos\theta}$ for an odd polynomial $g$ of
degree at most $K-1$, with $\sup_{[-1,1]}\abs{g}\le1$. Since
$q_K\rbra{x}=\int_0^x g\rbra{u}\,\mathrm d u$, the polynomial $q_K$ is real and
even, $\deg\rbra{q_K}\le K$, $q_K\rbra{0}=0$, and $\abs{q_K'}\le1$ on $[-1,1]$.
In particular, $\abs{q_K\rbra{x}}\le \abs{x}$ on this interval.

Translation by $t$ changes $\sgn\rbra{\cos\theta}$ by at most
$4\abs{t}$ in $L^1$. By \eqref{eq:jackson-bounds} and $x=\cos\theta$,
\begin{equation}\label{eq:app-derivative-integral}
 \int_{-1}^{1}\abs{g\rbra{x}-\sgn\rbra{x}}\,\mathrm d x
 \le \frac12\int_{-\pi}^{\pi}\abs{h_M\rbra{\theta}-\sgn\rbra{\cos\theta}}\,\mathrm d\theta
 \le \frac{2^6}{K}.
\end{equation}
By oddness of $g$ and $\sgn$, for $\abs{x}\leq1$,
\[
 \abs{q_K\rbra{x}-\abs{x}}
 =\abs*{\int_0^x\rbra{g\rbra{u}-\sgn\rbra{u}}\,\mathrm d u}
 \leq\frac12\int_{-1}^1\abs{g\rbra{u}-\sgn\rbra{u}}\,\mathrm d u
 \leq\frac{2^5}{K}.
\]
For $x\ne0$, the distance from $\arccos x$ to the nearest jump of
$\sgn\rbra{\cos\theta}$ is $\arcsin\abs{x}\ge\abs{x}$. Therefore
\eqref{eq:app-jackson-tail} gives
$\abs{q_K'\rbra{x}-\sgn\rbra{x}}\le\min\cbra{2,2^5/\rbra{M^3\abs{x}^3}}$.
Multiplying by $\abs{x}$ gives
\[
 \abs{x}\,\abs{q_K'\rbra{x}-\sgn\rbra{x}}
 \leq\min\cbra*{2\abs{x},\frac{2^5}{M^3\abs{x}^2}}
 \leq\frac{2^3}{M}\leq\frac{2^4}{K}.
\]
Together with \eqref{eq:app-derivative-integral}, this gives
\begin{equation}
\int_{-1}^1\abs{\sgn\rbra{x}-q_K'\rbra{x}}\,\mathrm d x\leq\frac{2^6}{K},
\qquad
\sup_{\abs{x}\leq1}\abs{x}\,\abs{\sgn\rbra{x}-q_K'\rbra{x}}\leq\frac{2^4}{K}.
\label{eq:absolute-errors}
\end{equation}

For the higher derivatives, the Chebyshev coefficient of $g$ at $T_\ell$ is
\[
 \frac{4\sin\rbra{\pi\ell/2}}{\pi\ell}
 \int_{-\pi}^{\pi}J_M\rbra{t}\cos\rbra{\ell t}\,\mathrm d t.
\]
Its magnitude is at most $4/\rbra{\pi\ell}$, and only odd $\ell<K$ occur. The interior bound
$\sup_{\abs{x}\leq1/2}\abs{T_\ell^{\rbra{r}}\rbra{x}}
\leq\rbra{2^2\ell}^r$ follows from \cref{lem:divided-differences} at repeated nodes.
Since $q_K^{\rbra{j}}=g^{\rbra{j-1}}$, for $2\leq j\leq K$,
\[
\sup_{\abs{x}\leq1/2}\abs{q_K^{\rbra{j}}\rbra{x}}
\leq 2^{2j}\sum_{\ell=j-1}^{K-1}\ell^{j-2}
\leq 2^{2j}K^{j-1}.
\]
These bounds prove \cref{lem:absolute} with $C=2^6$.
\end{proof}

\subsection{Trace approximation errors}

\begin{proof}[Proof of \cref{lem:trace-errors}]
% We show that $C=2^6$ suffices.
Set $e\rbra{t}=\abs{t}-q_{K,s}\rbra{t}$. This function is absolutely
continuous on $[-s,s]$, with
$e'\rbra{t}=\sgn\rbra{t}-q_{K,s}'\rbra{t}$ almost everywhere.
For $A=U,V$, let $N_A\rbra{t}=\tr\rbra{\mathbf1_{(-\infty,t]}\rbra{A}}$
count the eigenvalues of $A$ at most $t$, including multiplicities.
The spectral counting bound \cite[Theorem~A.43]{BS10} states that
\[
 \abs{N_U\rbra{t}-N_V\rbra{t}}\leq\rank\rbra{U-V}
 \qquad\text{for every }t\in\mathbb R.
\]
To make the integration-by-parts step explicit, write $n$ for the common
matrix size. Summing
$e\rbra{\lambda}=e\rbra{s}-\int_\lambda^s e'\rbra{t}\,\mathrm d t$
over the eigenvalues of $A$, all of which lie in $[-s,s]$, gives
\[
 \tr\rbra{e\rbra{A}}
 =n e\rbra{s}-\int_{-s}^s e'\rbra{t}N_A\rbra{t}\,\mathrm d t.
\]
The terms $n e\rbra{s}$ cancel when we subtract the identities for $U$ and $V$.
Since $E_{K,s}\rbra{A}=\tr\rbra{e\rbra{A}}$, it follows that
\begin{align*}
 \abs{E_{K,s}\rbra{U}-E_{K,s}\rbra{V}}
 &=\abs*{\int_{-s}^s e'\rbra{t}\rbra{N_V\rbra{t}-N_U\rbra{t}}\,\mathrm d t}\\
 &\leq\int_{-s}^s\abs{e'\rbra{t}}\,\abs{N_V\rbra{t}-N_U\rbra{t}}\,\mathrm d t\\
 &\leq\rank\rbra{U-V}\int_{-s}^{s}\abs{\sgn\rbra{t}-q_{K,s}'\rbra{t}}\,\mathrm d t\\
 &\leq\frac{2^6s}{K}\rank\rbra{U-V}.
\end{align*}
The last step uses $q_{K,s}'\rbra{t}=q_K'\rbra{t/s}$, the substitution
$t=sx$, and \eqref{eq:absolute-errors}.
\end{proof}

\begin{lemma}[Trace error under a congruence]
\label{lem:trace-congruence}
For an integer $K\geq2$, $s>0$, a Hermitian matrix $U$ with
$\norm{U}_\infty\leq s$, and $0\prec\Lambda\preceq\Id$,
\begin{equation}
\abs{E_{K,s}\rbra{\Lambda U\Lambda}-E_{K,s}\rbra{U}}
\leq\frac{2^5s}{K}\norm{\log\Lambda}_1.
\label{eq:trace-congruence-error}
\end{equation}
\end{lemma}

\begin{proof}
Set $U_t=\Lambda^t U\Lambda^t$ for $0\le t\le1$. Then $\norm{U_t}_\infty\le s$.
Invertible congruence makes that all zero eigenvalues
remain zero and the nonzero eigenvalues stay uniformly away from zero
along this compact path. Consequently $\tr\rbra{\abs{U_t}}$ is differentiable,
including when $U$ is singular. Cyclicity of trace gives
$\frac{\mathrm d}{\mathrm d t}E_{K,s}\rbra{U_t}
=2\tr\rbra{U_t\rbra{\sgn\rbra{U_t}-q_{K,s}'\rbra{U_t}}\log\Lambda}$.
By \eqref{eq:absolute-errors} and trace duality,
\begin{align*}
 \abs*{\frac{\mathrm d}{\mathrm d t}E_{K,s}\rbra{U_t}}
 &\leq2\norm{U_t\rbra{\sgn\rbra{U_t}-q_{K,s}'\rbra{U_t}}}_\infty
 \norm{\log\Lambda}_1\\
 &\leq\frac{2^5s}{K}\norm{\log\Lambda}_1.
\end{align*}
Integrating over $0\leq t\leq1$ proves \eqref{eq:trace-congruence-error}.
\end{proof}

\section{Proof of the multiscale polynomial bounds}
\label[appendix]{app:multiscale}

\subsection{Polynomial filters}
\label[appendix]{app:filter}

\begin{proof}[Proof of \cref{lem:filter}]
The assertions are immediate for $b\ge1/32$, when $R_b=1$.
Otherwise, let $m=\floor{\rbra{8b}^{-1/2}}\ge2$. For
$x=2\sin^2\rbra{\theta/2}\in[0,2]$, the definition of $T_m$ gives
$r_m\rbra{x}=\sin^2\rbra{m\theta/2}/\rbra{m^2\sin^2\rbra{\theta/2}}$.
Since $\abs{\sin\rbra{mt}}\le m\abs{\sin t}$, we have
$0\le r_m\rbra{x}\le\min\cbra{1,2/\rbra{m^2x}}$. Thus
$0\le R_b=2r_m-r_m^2\le1$, $R_b\rbra{0}=1$, and
$R_b\rbra{x}\leq\min\cbra{1,4/\rbra{m^2x}}$.
Hence $xR_b\rbra{x}^2\leq4/m^2\leq2^7b$, using $m\geq\rbra{32b}^{-1/2}$.
Also, $\deg\rbra{R_b}\le2\rbra{m-1}$.

Expanding $m^{-2}\abs{\sum_{a=0}^{m-1}e^{ia\theta}}^2$ yields
\begin{equation}\label{eq:app-fejer-expansion}
 r_m\rbra{x}=\frac{1}{m^2}
 \rbra*{m+2\sum_{\ell=1}^{m-1}\rbra{m-\ell}T_\ell\rbra{1-x}}.
\end{equation}
These coefficients are nonnegative and sum to one.
The bound $\sup_{[-1,1]}\abs{T_\ell'}\leq\ell^2$
\cite[Appendix~A.1]{CGW26} gives
$\sup_{[0,2]}\abs{r_m'}\leq\rbra{m-1}^2$. Since $m^2b\le1/8$, for
$0\le x\le b$ we obtain $0\le1-r_m\rbra{x}\le m^2x\le x/\rbra{8b}$ and
$1-R_b\rbra{x}=\rbra{1-r_m\rbra{x}}^2\le x^2/\rbra{64b^2}$.
Finally, $-\log\rbra{1-u}\leq u/\rbra{1-u}$ for $0\leq u<1$ gives
$-\log\rbra{R_b\rbra{x}}\leq x^2/\rbra{63b^2}\leq\rbra{x/b}^2$.
\end{proof}

\subsection{Approximation error and degree}
\label[appendix]{app:multiscale-approximation}

\begin{proof}[Approximation and degree bounds in \cref{prop:multiscale}]
Evaluate $Y_\ell$ at $\rbra{S_+,S_-}$ as in \eqref{eq:multiscale}.
By the order inequalities and \cref{lem:filter},
\begin{equation}\label{eq:app-terminal}
 \norm{Y_\ell}_\infty\le 2^{7-\ell},\qquad
 \norm{Y_J}_1\le d2^{7-J}\le\varepsilon/4.
\end{equation}
For the analysis, define
\begin{align*}
 P_\ell&=\mathbf1_{[2^{-\ell},\infty)}\rbra{S_+},&
 Q_\ell&=\Id-P_\ell,\\
 \widetilde R_\ell&=Q_\ell+P_\ell R_{2^{-\ell}}\rbra{S_+}P_\ell,&
 \widetilde Y_\ell&=\widetilde R_\ell S_-\widetilde R_\ell,
\end{align*}
with $\widetilde R_0=\Id$ and $\widetilde Y_0=S_-$.
Since $-\widetilde R_\ell S_+\widetilde R_\ell
\preceq\widetilde Y_\ell\preceq
\widetilde R_\ell S_+\widetilde R_\ell$, we have
$\norm{\widetilde Y_\ell}_\infty\le 2^{7-\ell}$.
Both $\widetilde R_{\ell-1}$ and $\widetilde R_\ell$ equal the
identity on $Q_\ell$, so
$\rank\rbra{\widetilde Y_{\ell-1}-\widetilde Y_\ell}\le2\rank\rbra{P_\ell}$.
Applying \cref{lem:trace-errors} at width
$2^{9-\ell}$ gives
\begin{equation}\label{eq:app-scale-rank}
 \abs{E_{K_\ell,2^{9-\ell}}\rbra{\widetilde Y_{\ell-1}}
    -E_{K_\ell,2^{9-\ell}}\rbra{\widetilde Y_\ell}}
 \le \frac{2^{16-\ell}\rank\rbra{P_\ell}}{K_\ell}.
\end{equation}

To compare these matrices with $Y_\ell$, set
$\Lambda_\ell=Q_\ell R_{2^{-\ell}}\rbra{S_+}Q_\ell+P_\ell$ and
$\Lambda_0=\Id$. Then $0\prec\Lambda_\ell\preceq\Id$,
$Y_\ell=\Lambda_\ell\widetilde Y_\ell\Lambda_\ell$, and
\cref{lem:filter} gives
$\norm{\log\Lambda_\ell}_1\le 2^{2\ell}\tr\rbra{S_+^2\mathbf1_{[0,2^{-\ell})}\rbra{S_+}}$.
Applying \cref{lem:trace-congruence} at both scales with the same
$q_{K_\ell,2^{9-\ell}}$ and using \eqref{eq:app-scale-rank} gives
\begin{align}
 \abs{E_{K_\ell,2^{9-\ell}}\rbra{Y_{\ell-1}}-E_{K_\ell,2^{9-\ell}}\rbra{Y_\ell}}
 &\le \frac{2^{16}w_\ell}{K_\ell},\label{eq:app-scale-error}\\
 w_\ell&=2^{-\ell}\rank\rbra{P_\ell}
 +2^\ell\tr\rbra[\big]{S_+^2\mathbf1_{[0,2^{-\ell})}\rbra{S_+}}\notag\\
 &\quad+2^{\ell-1}\tr\rbra[\big]{S_+^2\mathbf1_{[0,2^{1-\ell})}\rbra{S_+}},
 \label{eq:app-scale-weight}
\end{align}
where the last summand is omitted for $\ell=1$.

For each eigenvalue $\lambda\ge0$ of $S_+$, the dyadic scales satisfy
\[
 \sum_{\ell:2^{-\ell}\le\lambda}2^{-\ell}\le2\lambda,
 \qquad
 \sum_{\ell:2^{-\ell}>\lambda}2^\ell\lambda^2\le2\lambda.
\]
Consequently $\sum_\ell w_\ell\leq2^3\tr\rbra{S_+}\leq2^3$.
At the scales with $d2^{-\ell}\geq1$, we have $K_\ell\geq C/\varepsilon$.
Taking the constant in \eqref{eq:scales} sufficiently large gives
\[
 \sum_{\ell:d2^{-\ell}\geq1}\frac{2^{16}w_\ell}{K_\ell}
 \leq\frac{\varepsilon}{4}.
\]
At all other scales, $w_\ell\leq2^2d2^{-\ell}$, since
$\rank\rbra{P_\ell}\leq d$ and
$\tr\rbra{S_+^2\mathbf1_{[0,b)}\rbra{S_+}}\leq db^2$.
After increasing the constant in \eqref{eq:scales} if needed,
\[
 \sum_{\ell:d2^{-\ell}<1}\frac{2^{16}w_\ell}{K_\ell}
 \leq
 \frac{\varepsilon}{8}
 \sum_{\ell:d2^{-\ell}<1}\sqrt{d2^{-\ell}}
 \leq\frac{\varepsilon}{2}.
\]
The trace-norm differences telescope to
$\tr\rbra{\abs{S_-}}-\tr\rbra{\abs{Y_J}}$. Hence
\begin{align*}
 &\abs{F_{d,\varepsilon}\rbra{S_+,S_-}-\tr\rbra{\abs{S_-}}}\\
 &\quad\leq\norm{Y_J}_1+
 \sum_{\ell=1}^J\abs{E_{K_\ell,2^{9-\ell}}\rbra{Y_{\ell-1}}
 -E_{K_\ell,2^{9-\ell}}\rbra{Y_\ell}}\\
 &\quad\leq\frac{\varepsilon}{4}+\frac{\varepsilon}{4}
 +\frac{\varepsilon}{2}=\varepsilon.
\end{align*}

Since $2^9d/\varepsilon\leq2^J<2^{10}d/\varepsilon$,
rounding in the definition of $K_\ell$ gives
\begin{equation}\label{eq:app-scale-degrees}
 K_\ell\leq\frac{C}{\varepsilon}\min\cbra{1,\sqrt{d2^{-\ell}}}.
\end{equation}
Since $K_{\ell+1}\leq K_\ell$, \cref{lem:filter} gives
\begin{align*}
 \deg\rbra{F_{d,\varepsilon}}
 &\leq\max\cbra*{K_1,
 \max_{1\leq\ell\leq J}K_\ell\rbra{2\deg\rbra{R_{2^{-\ell}}}+1}}\\
 &\leq\max\cbra*{K_1,
 \max_{1\leq\ell\leq J}CK_\ell2^{\ell/2}}
 \leq C\sqrt d/\varepsilon.
\end{align*}
The definitions show that all coefficients depend only on $d,\varepsilon$.
\end{proof}

\subsection{Separated representations of divided differences}

For a polynomial $p$, let $p^{[a]}$ denote its divided difference of
order $a$, including repeated nodes. Equivalently, extend linearly the
identity
\begin{equation}\label{eq:app-monomial-dd}
 \rbra{x^k}^{[a]}\rbra{x_0,\ldots,x_a}
 =\sum_{r_0+\cdots+r_a=k-a}\prod_{i=0}^a x_i^{r_i},
\end{equation}
where the sum is zero for $a>k$.
A separated representation on $I^{a+1}$ is a finite sum of the form
\[
 \sum_\alpha c_\alpha\prod_{i=0}^a g_{\alpha,i}\rbra{x_i},
 \qquad\text{with bound}\qquad
 \sum_\alpha\abs{c_\alpha}\prod_i\sup_I\abs{g_{\alpha,i}}.
\]
This bound controls the associated ordered matrix products;
a scalar supremum bound alone does not suffice.

\begin{lemma}[Separated Chebyshev differences]\label{lem:divided-differences}
For integers $k\geq a\geq1$, the following functions admit separated
representations with these bounds:
\begin{enumerate}
\item $T_k^{[a]}$ on $[-1,1]^{a+1}$ has bound
$\rbra{2k^2}^a/\rbra{2a}!$;
\item $T_k^{[a]}$ on $[-1/2,1/2]^{a+1}$ has bound
$\rbra{2^2k}^a/a!$;
\item the order-$a$ divided difference of $x\mapsto T_k\rbra{1-x}$, multiplied by
$\sqrt{x_r}$, has bound $2^ak^{2a-1}/\rbra{2a-1}!$ on $[0,1]^{a+1}$,
for each $r\in\cbra{0,a}$.
\end{enumerate}
\end{lemma}

\begin{proof}
Let $U_\ell\rbra{\cos\theta}=\sin\rbra{\rbra{\ell+1}\theta}/\sin\theta$
\cite[Equation~(18.5.2)]{OLBC2010}, and set $v_0=1$, $v_\ell=2$ for $\ell\geq1$.
The generating functions \cite[Equations~(18.12.7) and (18.12.10)]{OLBC2010} give
\begin{equation}\label{eq:app-chebyshev-dd}
 T_k^{[a]}\rbra{x_0,\ldots,x_a}
 =2^{a-1}\sum_{\ell_0+\cdots+\ell_a=k-a}
 v_{\ell_0}T_{\ell_0}\rbra{x_0}\prod_{i=1}^a U_{\ell_i}\rbra{x_i}.
\end{equation}
Indeed, taking divided differences of
$\sum_{k\ge0}T_k\rbra{x}z^k=\frac12+\frac{1-z^2}{2\rbra{1-2xz+z^2}}$
gives
$2^{a-1}z^a\rbra{1-z^2}\prod_{i=0}^a\rbra{1-2x_i z+z^2}^{-1}$.
Absorb $1-z^2$ into the first factor and compare coefficients of $z^k$.

On $[-1,1]$, each factor in \eqref{eq:app-chebyshev-dd} attains
its supremum norm at $1$, and all coefficients are nonnegative.
Thus its representation bound equals
\[
 T_k^{[a]}\rbra{1,\ldots,1}
 =\frac{T_k^{\rbra{a}}\rbra{1}}{a!}
 \leq\frac{2^ak^{2a}}{\rbra{2a}!},
\]
where the inequality is from \cite[Appendix~A.1]{CGW26}.
On $[-1/2,1/2]$, instead use $\abs{U_\ell}\le2/\sqrt3$.
The $\binom ka$ compositions in \eqref{eq:app-chebyshev-dd} give the bound
\[
 2^{a-1}\sum_{\ell_0+\cdots+\ell_a=k-a}
 v_{\ell_0}\prod_{i=1}^a\sup_{\abs{x}\leq1/2}\abs{U_{\ell_i}\rbra{x}}
 \leq\rbra{4/\sqrt3}^a\binom ka
 \leq\frac{\rbra{4k}^a}{a!}.
\]

For $0<x\le1$, the definition of $U_\ell$ gives
$\sqrt{x}\abs{U_\ell\rbra{1-x}}\le1/\sqrt{2-x}\le1$.
By symmetry of divided differences, place the $T$ factor in
\eqref{eq:app-chebyshev-dd} at a node other than $x_r$.
The weighted $U$ factor has norm at most one. Bounding every unweighted
$U_\ell$ by $\ell+1$ gives a representation bound
\[
 2^{a-1}\rbra*{\binom{k+a-1}{2a-1}+\binom{k+a-2}{2a-1}}
 \leq\frac{2^ak^{2a-1}}{\rbra{2a-1}!}.
\]
Indeed, $\binom{k+a-2}{2a-1}\leq\binom{k+a-1}{2a-1}$ and
\[
 \binom{k+a-1}{2a-1}
 =\frac{k\prod_{r=1}^{a-1}\rbra{k^2-r^2}}{\rbra{2a-1}!}
 \leq\frac{k^{2a-1}}{\rbra{2a-1}!}.
\]
The affine substitution $x\mapsto1-x$ contributes only a sign
$\rbra{-1}^a$, and continuity includes $x=0$.
\end{proof}

For a Hermitian matrix $M=\sum_\lambda\lambda P_\lambda$ and an
arbitrary complex matrix $E$, the order-$a$ coefficient of $p\rbra{M+tE}$ is
\begin{equation}\label{eq:app-matrix-dd}
 \sum_{\lambda_0,\ldots,\lambda_a}
 p^{[a]}\rbra{\lambda_0,\ldots,\lambda_a}
 P_{\lambda_0}EP_{\lambda_1}\cdots EP_{\lambda_a}.
\end{equation}
For a monomial, this follows by inserting the spectral decomposition in
\[
 \sum_{r_0+\cdots+r_a=k-a}M^{r_0}EM^{r_1}\cdots EM^{r_a}.
\]
Extend by linearity. Each separated factor gives a product
$g_0\rbra{M}Eg_1\rbra{M}\cdots Eg_a\rbra{M}$ with the same bound on its known factors. When $M\succeq0$, a weight $\sqrt{\lambda_0}$ or
$\sqrt{\lambda_a}$ is multiplication by $M^{1/2}$ on the left or right.

\subsection{Taylor coefficients at a fixed scale}

For matrix-valued versions of \eqref{eq:variance-taylor-representation},
the bounds \eqref{eq:variance-taylor-bound} multiply under concatenation,
add under summation, and are unchanged by taking a trace.

\begin{lemma}[Taylor coefficients of a filtered polynomial]\label{lem:filtered}
Let $0<b\le1$, $K\ge2$ be an integer, and $2^8b\le s\le2^9b$.
Let $S_+,S_-$ be Hermitian matrices with $0\preceq S_+\preceq\Id$ and
$-S_+\preceq S_-\preceq S_+$. For arbitrary complex matrices $E_0,E_1$
of the same size, expand
\[
 \tr\rbra*{q_{K,s}\rbra*{R_b\rbra{S_+ +tE_0}\rbra{S_- +tE_1}R_b\rbra{S_+ +tE_0}}}
 =\sum_{j\ge0}t^j\Phi_j\rbra{E_0,E_1}.
\]
For each $j\ge1$, $\Phi_j$ has a representation of the form
\eqref{eq:variance-taylor-representation}, valid for all $E_0,E_1$,
whose bound in \eqref{eq:variance-taylor-bound} is at most
\begin{equation}\label{eq:app-filtered-bound}
 2^{10}Kb\,\frac{\rbra{2^{16}K^2/b}^j}{\rbra{2j}!}\qquad\rbra{j\ge1}.
\end{equation}
\end{lemma}

\begin{proof}
The coefficients in \eqref{eq:app-fejer-expansion} are nonnegative
and sum to one. The identity $2T_uT_v=T_{u+v}+T_{\abs{u-v}}$ shows that
the sum of the absolute Chebyshev coefficients is submultiplicative.
Thus this sum is at most
three for $R_b=2r_m-r_m^2$, and at most one in the constant branch.
Write
$R_b\rbra{S_+ +tE_0}=\sum_{a\ge0}t^aR_a\rbra{E_0}$.
By \cref{lem:divided-differences}, \eqref{eq:app-matrix-dd}, and
$\deg\rbra{R_b}\le b^{-1/2}$, the following bounds hold for $a\ge1$:
\begin{equation}\label{eq:app-filter-coefficients}
 \begin{array}{c|c}
 \text{coefficient}&\text{bound of its matrix-product representation}\\ \hline
 R_a\rbra{E_0}&2^{a+2}b^{-a}/\rbra{2a}!\\[2pt]
 R_a\rbra{E_0}S_+^{1/2}\text{ or }S_+^{1/2}R_a\rbra{E_0}
 &2^{a+4}b^{1/2-a}/\rbra{2a-1}!
 \end{array}
\end{equation}
At order zero, \cref{lem:filter} gives
\[
 \norm{R_0}_\infty\leq1,\qquad
 \max\cbra*{\norm{R_0S_+^{1/2}}_\infty,
 \norm{S_+^{1/2}R_0}_\infty}\leq2^4\sqrt b.
\]

Set $Y=R_b\rbra{S_+}S_- R_b\rbra{S_+}$ and expand the matrix inside $q_{K,s}$ as
$Y+\sum_{j\ge1}t^j\Delta_j$.
The coefficients are
\[
 \Delta_j=\sum_{a+a'=j}R_a\rbra{E_0}S_-R_{a'}\rbra{E_0}
 +\sum_{a+a'=j-1}R_a\rbra{E_0}E_1R_{a'}\rbra{E_0}.
\]
For the products $R_a\rbra{E_0}S_-R_{a'}\rbra{E_0}$, factor
$S_-=S_+^{1/2}J S_+^{1/2}$ with $\norm{J}_\infty\le1$ and apply the weighted
bounds in \eqref{eq:app-filter-coefficients}. Summing over $a+a'=j$ gives
\[
 2^8b\rbra{2/b}^j\sum_{a+a'=j}
 \frac{1}{\max\cbra{2a-1,0}!\max\cbra{2a'-1,0}!}
 \le \frac{2^7b\rbra{2^3/b}^j}{\rbra{2j-2}!}.
\]
The indices $a,a'\geq1$ give a binomial sum of degree $2j-2$;
the indices $a=0$ or $a'=0$ contribute $2/\rbra{2j-1}!$.
For the products $R_a\rbra{E_0}E_1R_{a'}\rbra{E_0}$,
sum over $a+a'=j-1$ to obtain the bound
\[
 2^4\rbra{2/b}^{j-1}\sum_{a+a'=j-1}\frac{1}{\rbra{2a}!\rbra{2a'}!}
 \le \frac{2b\rbra{2^3/b}^j}{\rbra{2j-2}!}.
\]
Consequently $\Delta_j$ has a matrix-product representation bounded by
\begin{equation}\label{eq:app-delta-coefficients}
 \frac{b\rbra{2^{11}/b}^j}{\rbra{2j-2}!}\qquad\rbra{j\ge1}.
\end{equation}
The two factors $S_+^{1/2}$ come from the factorization of $S_-$, so the
bound holds for arbitrary perturbations $E_0,E_1$.

Next, $\norm{Y}_\infty\le2^7b\le s/2$.
Since $q_K$ is even and $\abs{q_K}\leq1$ on $[-1,1]$,
the sum of its absolute Chebyshev coefficients is at most $K+1\leq2K$.
The interior estimate in \cref{lem:divided-differences}, after rescaling,
therefore bounds a separated representation of $q_{K,s}^{[a]}$ on
$[-s/2,s/2]^{a+1}$ by $2Ks\rbra{2^2K/s}^a/a!$.
Use \eqref{eq:app-matrix-dd} at $Y$ and substitute the ordered products
of the coefficients in \eqref{eq:app-delta-coefficients}.
The order-$j$ trace coefficient has a representation bounded by
\begin{equation}\label{eq:app-composition-sum}
 2^{10}Kb\rbra{2^{11}/b}^j\sum_{a=1}^{\min\cbra{K,j}}\frac{K^a}{a!}
 \sum_{\substack{\ell_1+\cdots+\ell_a=j\\\ell_i\ge1}}
 \prod_{i=1}^a\frac{1}{\rbra{2\ell_i-2}!}.
\end{equation}
The multinomial identity gives
\[
 \sum_{\substack{\ell_1+\cdots+\ell_a=j\\\ell_i\geq1}}
 \prod_{i=1}^a\frac1{\rbra{2\ell_i-2}!}
 \leq\sum_{\substack{r_1+\cdots+r_a=2j-2a\\r_i\geq0}}
 \prod_{i=1}^a\frac1{r_i!}
 =\frac{a^{2j-2a}}{\rbra{2j-2a}!}.
\]

We claim that, for all $j\ge1$,
\begin{equation}\label{eq:app-factorial-sum}
 \sum_{a=1}^{\min\cbra{K,j}}
 \frac{K^aa^{2j-2a}}{a!\rbra{2j-2a}!}
 \le\frac{\rbra{2^5K^2}^j}{\rbra{2j}!}.
\end{equation}
For $1\leq a\leq\min\cbra{K,j}$, use
$\rbra{2j}!/\rbra{2j-2a}!\leq\rbra{2j}^{2a}$ to obtain
\[
 \frac{\rbra{2j}!}{K^{2j}}
 \frac{K^aa^{2j-2a}}{a!\rbra{2j-2a}!}
 \leq
 \begin{cases}
 \dfrac{4^j j^{2j}}{K^j j!}\leq\rbra{4e}^j,&j\leq K,\\[6pt]
 \dfrac{\rbra{4j^2/K}^a}{a!}
 \leq\dfrac{\rbra{4j^2/K}^K}{K!}
 \leq\rbra{4e\rbra{j/K}^2}^K\leq\rbra{4e}^j,&j\geq K.
 \end{cases}
\]
For $j\leq K$, we used $a\leq j$ and $K^a/a!\leq K^j/j!$.
For $j\geq K$, the inequality
$\rbra{4e\rbra{j/K}^2}^K\leq\rbra{4e}^j$ follows because
$\rbra{\log\rbra{4e}+2\log\rbra{t}}/t$ decreases for $t\geq1$.
Summing over $a$ now gives
\[
 \frac{\rbra{2j}!}{K^{2j}}
 \sum_{a=1}^{\min\cbra{K,j}}
 \frac{K^aa^{2j-2a}}{a!\rbra{2j-2a}!}
 \leq j\rbra{4e}^j\leq\rbra{2^5}^j,
\]
which proves \eqref{eq:app-factorial-sum}. Substituting this into
\eqref{eq:app-composition-sum} proves the lemma.
The identities hold for arbitrary complex perturbations, with known
factors depending only on $S_+,S_-$.
\end{proof}

\subsection{Summation over scales}\label[appendix]{app:coefficients}

\begin{proof}[Coefficient bounds in \cref{prop:multiscale}]
At most two terms of $F_{d,\varepsilon}$ involve each $Y_\ell$.
Their widths are $2^{8-\ell}$ or $2^{9-\ell}$, so
\cref{lem:filtered} applies. Their outer degrees are at most $K_\ell$,
since $K_{\ell+1}\leq K_\ell$.
For $\ell=0$, use the identity filter and set $K_0=K_1$.
Adding the representations gives
\begin{equation}\label{eq:app-summed-coefficients}
 M_j\leq\frac{2^{11}}{\rbra{2j}!}
 \sum_{\ell=0}^J K_\ell2^{-\ell}
 \rbra*{2^{16+\ell}K_\ell^2}^j.
\end{equation}
By \eqref{eq:app-scale-degrees}, for $j\geq2$,
\begin{align*}
 M_j&\leq\frac{\rbra{Cd/\varepsilon^2}^j}{d\varepsilon\rbra{2j}!}
 \rbra*{\sum_{\ell:d2^{-\ell}\geq1}\rbra{d2^{-\ell}}^{1-j}
       +\sum_{\ell:d2^{-\ell}<1}\rbra{d2^{-\ell}}^{3/2}}\\
 &\hspace{2em}\leq\frac{\rbra{Cd/\varepsilon^2}^j}{d\varepsilon\rbra{2j}!}.
\end{align*}
Here we used the geometric-series bounds
\[
 \sum_{\ell:d2^{-\ell}\geq1}\rbra{d2^{-\ell}}^{1-j}\leq2,
 \qquad
 \sum_{\ell:d2^{-\ell}<1}\rbra{d2^{-\ell}}^{3/2}\leq2.
\]
For $j=1$, \eqref{eq:app-summed-coefficients} gives
\begin{align*}
 M_1&\leq2^{26}\sum_{\ell=0}^JK_\ell^3\\
 &\leq\frac C{\varepsilon^3}
 \sum_{\ell=0}^J\min\cbra{1,\rbra{d2^{-\ell}}^{3/2}}
 \leq\frac{C\log\rbra{d}}{\varepsilon^3}.
\end{align*}
These bounds apply to the coefficients of $F_{d,\varepsilon}$ itself, as
required by \cref{prop:taylor-variance}.
\end{proof}

\end{document}